\documentstyle[12pt]{article}
\def\nin{\noindent}
\def\Al{{\cal A}}
\def\Mod{{\cal M}}
\def\DiK{{\cal D}}
\def\F{{\cal F}}
\def\X{{\cal X}}
\def\R{{\cal R}}
\def\J{{\bf J}}
\def\JD{{\cal J}}
\def\Clif{{{\cal C}\ell}}
\def\CClif{{{\cal C}\ell}^{\bf C}}
\def\Id{{\cal I}}
\def\j{{\it i}}
\def\um{{\bf 1}}
\def\n{{\bf n}}
\def\h{{\bf h}}
\def\Proj{{\bf P}}
\def\RE{{\bf R}}
\def\MProj{{\cal P}}
\def\Hi{{\cal H}}
\def\L{{\cal L}}
\def\v{{\vartheta}}
\def\p{^{\,\prime}}
\def\tg{{\rm tg}}
\def\cotg{{\rm cotg}}
\def\ut{{\underline{\theta}}}

\def\dd{{\bf d}}
\def\du{{\bf d}_u}

\def\ua{\underline{\alpha}}
\def\ub{\underline{\beta}}
\def\uxi{{\underline{\xi}}}
\def\ueta{{\underline{\eta}}}
\def\pint{\iota}

\def\diff{{\cal C}^\infty(S^2,{\bf R})}
\def\dag{\dagger}
\def\Uda{\Omega_u}
\def\UDda{\Omega_D}
\renewcommand{\theequation}{\thesection.\arabic{equation}}
\begin{document}
%
\hfill {\bf UFRJ-IF-FPC-012/96}
\vskip 1cm
\begin{center}
{\Large \bf  The Connes-Lott Program on the Sphere}
\vskip 0,5cm
{\bf \large J.A.Mignaco${}^{(*)}$\footnote {Partially suported by CNPq, Brasil, E-mail: mignaco @if.ufrj.br },
C.Sigaud${}^{(*)}$\footnote{E-mail: sigaud@if.ufrj.br},\\
A.R.da Silva${}^{(**)}$\footnote{E-mail: beto@lpim.ufrj.br}
and F.J.Vanhecke${}^{(*)}$\footnote{E-mail: vanhecke@if.ufrj.br }}
\vskip 0.5cm
${}^{(*)}${\it  Instituto de F\'\i sica}, ${}^{(**)}${\it Instituto de Matem\'atica,}\\
{\it UFRJ, Ilha do Fund\~ao, Rio de Janeiro, Brasil.}\\
\end{center}
\begin{abstract}
We describe the classical Schwinger model as a study of 
 the projective modules over the algebra of complex-valued functions on the
sphere. On these modules, classified by $\pi_2(S^2)$, we construct hermitian 
connections with values in the universal differential envelope which leads 
us to the Schwinger model on the sphere. The Connes-Lott program is then 
applied using the Hilbert space of complexified inhomogeneous forms with its 
Atiyah-K\"{a}hler structure. This Hilbert space splits  in two minimal 
left ideals of the Clifford algebra preserved by the Dirac-K\"{a}hler operator 
${\bf D}=\j(\dd-\delta)$. The induced representation of the universal 
differential envelope, in order to recover its differential structure, is divided 
by the unwanted differential ideal and the obtained quotient is the  usual
complexified de Rham exterior algebra over the sphere with Clifford 
action on the "spinors" of the Hilbert space. The subsequent steps of the 
Connes-Lott program allow to define a matter action, and the field action
is obtained using the Dixmier trace which reduces to the integral of 
the curvature squared.
\end{abstract}
\newpage
\section{Introduction}
\setcounter{equation}{0}
Since Dirac's seminal paper \cite{Dirac}, the static magnetic
monopole has been of considerable interest, not as much as a
physically realised system but mainly as an example of a theory with 
nontrivial topological features. It is a curious coincidence that
Dirac's paper appeared in the same year as Hopf's paper
\cite{Hopf} on circle bundles over the sphere.
The geometrical fibre bundle approach appeared in the seventies by
Wu and Yang \cite{W-Y} in a language more accessible to physicists
and by Greub and Petry \cite{G-P} in a more formal way.
Another approach using the differential characters of Cheeger and 
Simons \cite{Ch-S} was made by Coquereaux \cite{Coq1}.
The monopole is described by its magnetic field defined on 
$\RE^3\backslash\{O\}$, homeomorphic to $S^2\times \RE_+$, of which 
the sphere $S^2$ is a deformation retract.
The non-trivial topological features are thus common to both 
$\RE^3\backslash\{O\}$ and $S^2$ and for simplicity reasons
we restrict ourselves to the case of the sphere on which we include spinor fields.
This actually means that we switched from the Dirac monopole in three
dimensions to the Schwinger model on the sphere.

Here we examine this model in the light of 
Serre-Swan's theorem on the correspondence between the sections
of (complex) vector bundles over the sphere $S^2$ and 
the projective modules over the algebra of smooth complex-valued functions 
on the sphere $\Al=\diff\otimes{\bf C}$.
We then apply the Connes-Lott program\footnote{We refer to the 
original Connes-Lott paper \cite{Con-L}, to the review article
\cite{V-GB} and to Connes' book \cite{Con1}.}
with a suitable Hilbert space and Dirac operator and 
obtain the results similar to those of the (classical)
Schwinger model on the sphere.\cite{Jaye}

The main interest of our work consists in an explicitly worked out example
of the Connes-Lott scheme applied to a relatively simple system.
Since the approach is entirely algebraic, it is feasible to generalise to 
the tensor product ${\cal C}^\infty(M,\RE)\,\otimes\,\Al_F$,
where $\Al_F$ is a finite, involutive algebra,
not necessarily commutative, e.g. the "standard model" choice
${\bf C}\oplus{\bf H}\oplus{\bf M}_3({\bf C})$.

In section {\bf 2} we define our notation for
the relevant differential geometry on the sphere.
The classification in topological sectors of the hermitian finite 
projective modules  $\Mod_\Proj$
over $\Al$ is given in section {\bf 3}.
The hermitian connections $\nabla_\Proj$ with values in the universal differential
envelope $\Omega^\bullet\bigl(\Al\bigr)$ are also constructed.
In section {\bf 4} we make our choice of a Hilbert space $\Hi$ 
and a Dirac operator {\bf D}. Instead of the usual square integrable spinors on the sphere 
and the usual Dirac operator associated to the 
Levi-Civita connection, we choose the space of sections of the Atiyah-K\"ahler bundle 
with its Clifford module structure 
and with ${\bf D}=i({\bf d}-\delta)$ as Dirac operator.
We do not discuss here the possible physical interpretation of these
"K\"ahler spinors" which remains controversial and for which we refer
to \cite{Graf},\cite{B-T} and \cite{Cru}. 
The representation of the algebra in $\Hi$, together with the Dirac 
operator {\bf D} induces a representation of the universal 
envelope $\Omega^\bullet\bigl(\Al\bigr)$ in $\Hi$. 
However, this representation is not a differential one 
and an unwanted ideal has to be quotiented out. 
The elimination of this so-called "junk" is done in the standard manner \cite{Con1} and 
yields a representation in $\Hi$ of the connection with values in the de Rham exterior algebra. 
Finally, in section {\bf 5} the projective modules $\Mod_\Proj$ are tensored over $\Al$ with
the Hilbert space $\Hi$ allowing to construct a covariant Dirac operator
$\DiK(\nabla_\Proj)$ in $\Mod_\Proj\otimes_a\Hi$, used in the
matter action.
The Yang-Mills functional, in 
each topological sector, is obtained by the Dixmier trace of the square of the corresponding 
curvature operator.
Some conclusions are drawn and further prospects are presented in section {\bf 6}.\\
The appendix contains a miscelanea of formulae useful in explicit calculations.
\section{Differential Geometry on the Sphere $S^2$}
\setcounter{equation}{0}
The sphere of radius $r$ is defined as 
$S^2=\bigl\{p=(x,y,z)\in {\bf R}^3\vert\; x^2+y^2+z^2=r^2\bigr\}$
and the stereographic projection  
on the equatorial plane is given in the austral chart $H_A=\bigl\{p\in S^2\vert z<r\bigr\}$ by: 
\begin{equation}
\label{austral}
\phi_A:H_A\rightarrow {\bf R}^2:(x,y,z)\rightarrow (\xi,\eta)\;,\;\mbox{where}\;
\xi={x\over r-z}\;,\;\eta={y\over r-z}\;,
\end{equation}
with inverse
\[
{\phi_A}^{-1}:{\bf R}^2\rightarrow H_A:(\xi,\eta)\rightarrow (x,y,z)\;,\;\mbox{where}
\]
\[
{x\over r}={2\xi\over 1+\xi^2+\eta^2}\,,\,
 {y\over r}={2\eta\over 1+\xi^2+\eta^2}\,,\,
{z\over r}={\xi^2+\eta^2-1\over 1+\xi^2+\eta^2}\;.
\]
In the boreal chart $H_B=\bigl\{p\in S^2\vert -r<z\bigr\}$ one has
\begin{equation}
\label{boreal}
\phi_B:H_B\rightarrow {\bf R}^2:(x,y,z)\rightarrow (\xi\p,\eta\p)\;,\;\mbox{where}\;
\xi\p={x\over r+z}\;,\;\eta\p={y\over r+z}\,,
\end{equation}
with inverse
\[
{\phi_B}^{-1}:{\bf R}^2\rightarrow H_B:(\xi\p,\eta\p)\rightarrow (x,y,z)\;,\;\mbox{where} 
\]
\[
{x\over r}={2\xi\p\over 1+{\xi\p}^2+{\eta\p}^2}\,,\, 
{y\over r}={2\eta\p\over 1+{\xi\p}^2+{\eta\p}^2}\,,\,
  {z\over r}={1-{\xi\p}^2-{\eta\p}^2\over 1+{\xi\p}^2+{\eta\p}^2}\;.
\]
It is useful to introduce the complex coordinates
$\zeta_A=\xi+\j \eta$ with its complex conjugate $\zeta^c_A=\xi-\j \eta$ in $H_A$ and
$\zeta_B =\xi\p -\j \eta\p\;,\;\zeta^c_B=\xi\p+\j \eta\p$ in $H_B$.\\
In the overlap $H_A\bigcap H_B=\bigl\{(x,y.z)\in S^2\vert -r<z<r\bigr\}$,
both coordinates are related by $\zeta_A\,\zeta_B = 1$, displaying the complex structure of 
$S^2\simeq{\bf CP}_1$.\\
Spherical ccordinates are defined on this overlap by :
\[
x=r\sin\theta\cos\varphi\;,\;y=r\sin\theta\sin\varphi\;,\;
z=r\cos\theta\,,
\]
related to the coordinates above by
\begin{equation}
\label{espherical}
\zeta_A=\cotg(\theta/2)\,\exp(+\j\varphi)\;,\;
\zeta_B=\tg(\theta/2)\,\exp(-\j\varphi)\;.
\end{equation}
Note that the signs are chosen such that the orientation defined by $(\xi,\eta)$ corresponds to a positive 
inward oriented normal on $S^2$ as imbedded in ${\bf R}^3$ and is opposed to that of $(\xi\p,\eta\p)$ 
and $(\theta,\varphi)$.\\
The metric on the sphere $S^2$ is obtained as the pull-back of the 
Euclidean metric on ${\bf R}^3\;:\;g_E=dx\otimes dx+dy\otimes dy+dz\otimes dz\;.$
In $H_A$ it is written as : 
\begin{equation}
\label{metricA1}
{\bf g}_{\vert A}= \ut^\xi\otimes\ut^\xi + \ut^\eta\otimes\ut^\eta\;,
\end{equation}
with the Zweibein 
$$\ut^\xi={2\over f_A}\dd\xi\;\hbox{and}\;
\ut^\eta={2\over f_A}\dd\eta\;,\;\hbox{where}\;f_A=1+\vert\zeta_A\vert^2\,.$$
In terms of $\ut_A=\ut^\xi+\j\ut^\eta={2\over f_A}\dd\zeta_A$ 
and its conjugate $\ut_A^c$, the metric reads :
\begin{equation}
\label{metricA2}
{\bf g}_{\vert A}= 
{1\over2}\bigl(\ut_A^c\otimes\ut_A + \ut_A\otimes\ut_A^c\bigr)\;.
\end{equation}
Similarly in  $H_B$, the Zweibein is given by
$$\ut^{\xi\prime}={2\over f_B}\dd\xi\p\;\hbox{and}\;
\ut^{\eta\prime}={2\over f_B}\dd\eta\p\;,\hbox{where}\;f_B
=1+\vert\zeta_B\vert^2\,,$$
or by 
$\ut_B=\ut^{\xi\prime}-\j\ut^{\eta\prime}={2\over f_B}\dd\zeta_B$  and 
$\ut^c_B$, so that the metric reads :
\begin{equation}
\label{metricB}
{\bf g}_{\vert B} = \ut^{\xi\prime}\otimes\ut^{\xi\prime} + 
\ut^{\eta\prime}\otimes\ut^{\eta\prime}
= {1\over2}\bigl(\ut_B^c\otimes\ut_B 
+ \ut_B\otimes\ut_B^c\bigr)\,.
\end{equation}
In the overlap $H_A\bigcap H_B$, one has the relation
\begin{equation}
\label{overlap1}
\ut_A=-{\zeta_A^2\over\vert\zeta_A\vert^2}\ut_B\;,
\;\mbox{and}\;
\ut^c_A=-{{\zeta^c_A}^2\over\vert\zeta_A\vert^2}\ut^c_B\;.
\end{equation}
In this overlap,  spherical coordinates can also be used and one has
\begin{eqnarray*}
f_A^{-1}=\sin^2(\theta/2)&,& f_B^{-1}=\cos^2(\theta/2)\;;\\
\ut_A=-\exp(\j\varphi)
\Bigl(\dd\theta-\j\sin \theta\,\dd\varphi\Bigr)&,&
\ut_B=\exp(-\j\varphi)\bigl(\dd\theta-\j\sin \theta\,\dd\varphi\bigr)\;.
\end{eqnarray*}
In $H_A$ the structure functions of the Zweibein field are given by :
\[
\dd\ut^\xi=\eta\,\ut^\xi\wedge\ut^\eta\;\mbox{and}\;
\dd\ut^\eta=\xi\,\ut^\eta\wedge\ut^\xi\;\mbox{or by}\]
\begin{equation}
\label{zweibein} 
\dd\ut_A=1/2\,\zeta_A\,\ut_A\wedge\ut_A^c\;,
\end{equation}
and the Levi-Civita connection reads : 
\[
\nabla\ut^\xi=-\Bigl(\xi\ut^\eta-\eta\ut^\xi\Bigr)\otimes\ut^\eta\;\mbox{and}\;
\nabla\ut^\eta=-\Bigl(\eta\ut^\xi-\xi\ut^\eta\Bigr)\otimes\ut^\xi\;\mbox{or by}\]
\begin{equation}\label{connection}
\nabla\ut_A=-1/2\Bigl(\zeta_A\,\ut_A^c-\zeta_A^c\ut_A\Bigr)\otimes\ut_A\;.
\end{equation}
Similar expressions hold in $H_B$ and the relation with 
(\ref{zweibein}) and (\ref{connection}), in the intersection $H_A\bigcap H_B$,
is easily worked out.

\nin
The oriented area element $\omega$ on the sphere is given, respectively in $H_A$ and $H_B$  by : 
\begin{eqnarray}
\omega_{\vert A}&=&\tau:=\ut^\xi\wedge\ut^\eta=
{\j\over2}\ut_A\wedge\ut_A^c\nonumber\\
\omega_{\vert B}&=&-\tau\p:=-\ut^{\xi\p}\wedge\ut^{\eta\p}=
{\j\over2}\ut_B\wedge\ut_B^c\,.\label{volume}
\end{eqnarray}
and in the overlap $H_A\bigcap H_B$, one has 
$$\omega_{\vert A}=\omega_{\vert B}=-\dd\theta\wedge\sin\theta\dd\varphi\;.$$
On the sphere, a local basis of the space of 
differential forms $\F^\bullet(M)$, is given in $H_A$ by 
$$\bigl\{1,\ut^\xi,\ut^\eta,\tau\bigr\}\;.$$
The only non zero $\diff$-valued scalar products of these basis vectors, as defined 
in (\ref{scalarproduct}), are:
$$
g^{-1}\bigl(1,1\bigr) =
g^{-1}\bigl(\ut^\xi,\ut^\xi\bigr)=
g^{-1}\bigl(\ut^\eta,\ut^\eta\bigr)=
g^{-1}\bigl(\tau,\tau\bigr)=1\;.$$
The multiplication table of the real Clifford algebra 
$\Clif(S^2)$, defined in (\ref{AtKae}) of the appendix, is given in table (\ref{eerste}).
\begin{table}[h]
\caption{Clifford multiplication}
\label{eerste}
\begin{center}
\begin{tabular}{|c||r r r r|}\hline
$\vee$&{\bf 1}&$\ut^\xi$&$\ut^\eta$&$\tau$\\ \hline\hline
{\bf 1}&{\bf 1}&$\ut^\xi$&$\ut^\eta$&$\tau$\\
$\ut^\xi$&$\ut^\xi$&{\bf 1}&$\tau$&$\ut^\eta$\\
$\ut^\eta$&$\ut^\eta$&-$\tau$&{\bf 1}&-$\ut^\xi$\\
$\tau$&$\tau$&-$\ut^\eta$&$\ut^\xi$&-{\bf 1}\\
\hline
\end{tabular}
\end{center}
\end{table}

\nin
Besides the trivial idempotent \um , 
the primitive idempotents are of the form
$P=1/2\bigl(\um+b_\xi\ut^\xi+b_\eta\ut^\eta+c\tau\bigr),\,$
with $b_\xi^2+b_\eta^2=1+c^2\;,\;b_\xi,b_\eta,c\,\in\,\diff\;.$\\
The Dirac-K\H{a}hler operators, $\DiK=\j(\dd-\delta)$ 
and $\DiK^D=\j(\dd^D-\delta^D)$,
defined in  (\ref{DiKae1}) and (\ref{DiKae2}), 
conserve the left-, $\Id^E_P=\Clif(S^2)\vee P$, 
and right-ideals, $\Id^D_P=P\vee\Clif(S^2)$,
if and only if $\nabla_XP=0$ for any vector field $X$ on the sphere.
This happens if $b_\xi=b_\eta=0$ and $1+c^2=0$, so that there is
no real solution. Thus, although the real representation of $\Clif(S^2)$ is reducible, 
the invariant subspaces are not conserved by the Dirac-K\H{a}hler operators.
However, in the complexified Clifford algebra
$\CClif(S^2)$, there is a solution, namely 
$P_\pm=1/2(1\pm\j\omega)$, so that 
the minimal left-, $\Id^E_\pm=\CClif(S^2)\vee P_\pm$, and 
right-ideals, $\Id^D_\pm=P_\pm\vee\CClif(S^2)$, are 
conserved by $\DiK$, respectively $\DiK^D$.\\
The complexified Clifford algebra $\CClif(S^2)$ can then 
be decomposed in a sum of minimal left ideals:
$\CClif(S^2)=\Id^E_+\,+\,\Id^E_-$,
providing two equivalent representations of 
$\CClif(S^2)$ on these ideals $\Id^E_\pm$.
To write down explicit matrices for these representations, one
chooses a local basis of $\Id^E_\pm$ in each chart, \\
e.g. in $H_A$, one takes $\bigl(P_\pm^A\;\;Q_\pm^A\bigr)\;,$ where
$Q_\pm^A=\ut^\xi\vee P_\pm^A\;.$
\begin{eqnarray}
P_+^A={1\over 2}(1+\j\tau)\;,\;
&Q_+^A={1\over 2}(\ut^\xi+\j\ut^\eta)=
&{1\over f_A}\dd\zeta_A\;,\\
P_-^A={1\over 2}(1-\j\tau)\;,\;
&Q_-^A={1\over 2}(\ut^\xi-\j\ut^\eta)=
&{1\over f_A}\dd\zeta_A^c\;.
\end{eqnarray}
In $H_B$, one has
$\bigl(P_\pm^B\;\;Q_\pm^B\bigr)\;,$ where
$Q_\pm^B=\ut^{\xi\p}\vee P_\pm^B\;$ and 
\begin{eqnarray}
P_+^B={1\over 2}(1-\j\tau\p)\;,\;
&Q_+^B={1\over 2}(\ut^{\xi\p}-\j\ut^{\eta\p})=
&{1\over f_B}\dd\zeta_B\;,\\
P_-^B={1\over 2}(1+\j\tau\p)\;,\;
&Q_-^B={1\over 2}(\ut^{\xi\p}+\j\ut^{\eta\p})=
&{1\over f_B}\dd\zeta_B^c\;,
\end{eqnarray}
with the relation in $H_A\bigcap H_B$ given by  :
\begin{equation}
\label{gaugetransform1}
\bigl(P_\pm^A\;\;Q_\pm^A\bigr)=
\bigl(P_\pm^B\;\;Q_\pm^B\bigr)\;{\cal T}_\pm^{BA}\;,\;
\end{equation}
where
$${\cal T}_+^{BA}= 
\left(
\begin{array}{cc}
1 & 0\\
0 & -({\zeta_A\over\vert\zeta_A\vert})^2
\end{array}
\right)\;,\;
{\cal T}_-^{BA}= 
\left(
\begin{array}{cc}
1 & 0\\
0 & -({\zeta_A^c\over\vert\zeta_A\vert})^2
\end{array}
\right)\;.$$
A general inhomogeneous form is now written as 
$\Psi=\Psi^{(+)}+\Psi^{(-)}$, with
\begin{eqnarray}
{\Psi^{(\pm)}}_{\vert A}&=&\bigl(P_\pm^A\;\;Q_\pm^A\bigr)\;
\left(
\begin{array}{c}
\psi^{A,(0)}_\pm\\
\psi^{A,(1)}_\pm
\end{array}
\right)\;,\nonumber\\
{\Psi^{(\pm)}}_{\vert B}&=&\bigl(P_\pm^B\;\;Q_\pm^B\bigr)\;
\left(
\begin{array}{c}
\psi^{B,(0)}_\pm\\
\psi^{B,(1)}_\pm
\end{array}
\right)\;\label{inhomform}
\end{eqnarray}
and, in $H_A\bigcap H_B$, one has the gauge transformation :
\begin{equation}
\label{gaugetransform2}
\left(
\begin{array}{c}
\psi^{B,(0)}_\pm\\
\psi^{B,(1)}_\pm
\end{array}
\right)= 
{\cal T}_\pm^{BA}
\left(
\begin{array}{c}
\psi^{A,(0)}_\pm\\
\psi^{A,(1)}_\pm
\end{array}
\right)\;.
\end{equation}
An element $\Phi$ of the Clifford algebra is then represented in each
ideal $\Id^E_\pm$, locally by the matrices $\R^A_\pm(\Phi)$,
$\R^B_\pm(\Phi)$ : 
\begin{eqnarray}
\Phi\vee\bigl(P^A_\pm\;\;Q^A_\pm\bigr)&=&
\bigl(P^A_\pm\;\;Q^A_\pm\bigr)\;\R^A_\pm(\Phi)\;,\label{repclif1}\\
\Phi\vee\bigl(P^B_\pm\;\;Q^B_\pm\bigr)&=&
\bigl(P^B_\pm\;\;Q^B_\pm\bigr)\;\R^B_\pm(\Phi)\;.\label{repclif2}
\end{eqnarray}
\newpage
\nin
Explicit matrices are obtained from table (\ref{voorstelling}).
\begin{table}[h]
\caption{Representation of the Clifford algebra}
\label{voorstelling}
\begin{center}
\begin{tabular}{|c||r r r r|}\hline
$\vee$&$P^A_+$&$Q^A_+$&$P^A_-$&$Q^A_-$\\ \hline\hline
{\bf 1}&$P^A_+$&$Q^A_+$&$P^A_-$&$Q^A_-$\\
$\ut^\xi$&$Q^A_+$&$P^A_+$&$Q^A_-$&$P^A_-$\\
$\ut^\eta$&-$\j Q^A_+$&$\j P^A_+$&$\j Q^A_-$&-$\j P^A_-$\\
$\tau$&-$\j P^A_+$&$\j Q^A_+$&$\j P^A_-$&-$\j Q^A_-$\\
\hline
\end{tabular}
\end{center}
\end{table}
\begin{eqnarray}
\R^A_\pm(\um)=\left(
\begin{array}{cc}
1 & 0\\
0 & 1
\end{array}
\right)
\;&,&\;
\R^A_\pm(\tau)=\left(
\begin{array}{cc}
\mp\j & 0\\
0 & \pm\j
\end{array}
\right)\;,\nonumber\\
\R^A_\pm(\ut^\xi)=\left(
\begin{array}{cc}
0 & 1\\
1 & 0
\end{array}
\right)
\;&,&\;
\R^A_\pm(\ut^\eta)=\left(
\begin{array}{cc}
0 & \pm\j \\
\mp\j & 0
\end{array}
\right)\;.
\label{repclif3}
\end{eqnarray}
In the overlap region $H_A\bigcap H_B$, one has:
\begin{equation}
\label{gaugetransform3}
\R^B_\pm(\Phi)=
{\cal T}_\pm^{BA}\;\;\R^A_\pm(\Phi)\;\;
\Bigl({\cal T}_\pm^{BA}\Bigr)^{-1}\;.
\end{equation} 
In $\CClif(S^2)$, the complex conjugation yields
$(P^A_\pm)^c=P^A_\mp\;,\;(Q^A_\pm)^c=Q^A_\mp$ so that the basis 
$\bigl(P^A_\pm\;\;Q^A_\pm\bigr)$ is orthonormal (see table (\ref{ortogon}))
with respect to the sesquilinear form  
$h^{-1}(\Psi,\Phi)=g^{-1}(\Psi^c,\Phi)$, given in (\ref{sesquilin}).

\begin{table}[h]
\caption{ Orthonormality relations}
\label{ortogon}
\begin{center}
\begin{tabular}{|c||r r r r|}\hline
$h^{-1}$&$P^A_+$&$Q^A_+$&$P^A_-$&$Q^A_-$\\ \hline\hline
$P^A_+$ &$1/2$ &$0$   &$0$   &$0$ \\
$Q^A_+$ &$0$   &$1/2$ &$0$   &$0$ \\
$P^A_-$ &$0$   &$0$   &$1/2$ &$0$ \\
$Q^A_-$ &$0$   &$0$   &$0$   &$1/2$\\
\hline
\end{tabular}
\end{center}
\end{table}

\nin
This hermitian scalar product with values 
in $\diff^{\bf C}$ is given by :
\begin{equation}
\label{sesquilins2}
h^{-1}(\Psi,\Phi)=
h^{-1}(\Psi^{(+)},\Phi^{(+)})+
h^{-1}(\Psi^{(-)},\Phi^{(-)})\,,
\end{equation}
where, in chart $H_A$ say, 
\begin{equation}
\label{localsesquilin}
h^{-1}(\Psi^{(\pm)}_{\vert A},\Phi^{(\pm)}_{\vert A})=
{1\over 2}\,
\Bigl(
(\psi^{A,(0)}_\pm)^c\phi_\pm^{A,(0)}\,+
(\psi^{A,(1)}_\pm)^c\phi_\pm^{A,(1)}
\Bigr)\;.
\end{equation}
Integrating over the sphere yields, in each $\Id^E_\pm$,
a hermitian scalar product, as in (\ref{hermitprod}), with complex values :
\begin{equation}
\label{hermitprods2}
\langle\Psi^{(\pm)}\;\vert\;\Phi^{(\pm)}\rangle=
\int_{S^2}\,h^{-1}\bigl(\Psi^{(\pm)},\Phi^{(\pm)}\bigr)\,\omega\;.
\end{equation}
Each $\Id^E_\pm$ is then completed with this product to form 
a Hilbert space $\Hi(P_\pm)$ and the total Hilbert space is the direct
orthogonal sum $\Hi(P_+) \oplus \Hi(P_-)$.
\nin
The Dirac-K\H{a}hler operator conserves this splitting and 
reads : 
\begin{equation}
\label{Dikaes2}
\DiK\Psi=\DiK_{(+)}\Psi^{(+)}\,+\,\DiK_{(-)}\Psi^{(-)}\,
\end{equation}
with, on each $\Hi(P_\pm)$ :
\begin{equation}
\label{locDiKaes2}
\DiK^A_{(\pm)}\Psi_{\vert A}^{(\pm)}=
\bigl(P^A_\pm\;\;Q^A_\pm\bigr)\,
\left(
\begin{array}{cc}
0 & \DiK^{A,(1)}_{(\pm)}\\
\DiK^{A,(0)}_{(\pm)} & 0
\end{array}
\right)
\left(
\begin{array}{c}
\psi^{A,(0)}_\pm\\
\psi^{A,(1)}_\pm
\end{array}
\right)\;.
\end{equation}
Each Hilbert space $\Hi(P_\pm)$ is ${\bf Z}_2$ graded by the 
parity of the corresponding differential form with grading 
operator $\pm\j\omega$ represented by the matrix 
\begin{equation}
\label{grading}
\left(\begin{array}{cc}1 & 0\\ 0 & -1 \end{array}\right)\,.
\end{equation}
With respect to this grading, $\Hi(P_\pm)$ is further decomposed as \\ $\Hi(P_\pm)=\Hi^{(0)}(P_\pm)\oplus\Hi^{(1)}(P_\pm)$ and the 
Dirac-K\H{a}hler operator is odd.\\
The operators $\DiK^{(0)}_{(\pm)}:\Hi^{(0)}(P_\pm)\rightarrow\Hi^{(1)}(P_\pm)$ 
and $\DiK^{(1)}_{(\pm)}:\Hi^{(1)}(P_\pm)\rightarrow\Hi^{(0)}(P_\pm)\;$
are formally adjoint in the sense that
$$
\langle\Psi_\pm^{(1)}\vert\DiK^{(0)}_{(\pm)}\Phi_\pm^{(0)}\rangle
-
\langle\DiK^{(1)}_{(\pm)}\Psi_\pm^{(1)}\vert\Phi_\pm^{(0)}\rangle=0\;.$$
It is easier to use the complex coordinates $\zeta_A$ and $\zeta_A^c$ 
to write explicit expressions for $\DiK^A_{(\pm)}$ :
\begin{eqnarray}
\DiK^A_{(+)}&=&
\j\,\left(
\begin{array}{cc}
0 & f_A\,{\partial\over\partial\zeta_A^c}\,-\,\zeta_A\\
f_A\,{\partial\over\partial\zeta_A}& 0 
\end{array}
\right)\label{locDis21}\\
\DiK^A_{(-)}&=&
\j\,\left(
\begin{array}{cc}
0 & f_A\,{\partial\over\partial\zeta_A}\,-\,\zeta_A^c\\
f_A\,{\partial\over\partial\zeta_A^c}& 0 
\end{array}
\right)\label{locDis22}
\end{eqnarray}
The index of the Dirac-K\H{a}hler operator, restricted to the even forms, is :
\begin{eqnarray}
Index\Bigl(\DiK^{(0)}_{(\pm)}\Bigr)
&=&
\hbox{dim}(Ker\,\DiK^{(0)}_{(\pm)})-\hbox{dim}(Coker\,\DiK^{0)}_{(\pm)})\nonumber\\
&=&
\hbox{dim}(Ker\,\DiK^{(0)}_{(\pm)})-\hbox{dim}(Ker\,\DiK^{(1)}_{(\pm)}).\label{index}
\end{eqnarray}
\nin
Now, $Ker\,\DiK^{(0)}_{(+)}$ is given by the functions $\psi$ such that
$f{\partial\over\partial\zeta_A}\psi=0$, i.e. the anti-holomorphic
functions on the sphere, and these are the constants, so that
$\hbox{dim}(Ker\,\DiK^{(0)}_{(+)})=1$. 
On the other hand, $Ker\,\DiK^{(1)}_{(+)}$ is given by  $\phi$ such that
$\Bigl(f\,{\partial\over\partial\zeta_A^c}\,-\,\zeta_A\Bigr)\phi=0$.
The substitution $\phi=f\,\gamma$ yields 
$f^2{\partial\over\partial\zeta_A^c}\gamma=0$, so that the solutions are 
$\phi=f\times \hbox{constant}$. 
These, however, are not integrable
over the sphere, so that
$\hbox{dim}(Ker\,\DiK^{(1)}_{(+)})=0$ and $\hbox{Index}\,\DiK^{(+,0)}=1-0$. \\
The same argument applies to $\DiK^{(0)}_{(-)}$ and the total index is
\begin{equation}
\label{euler}
Index\Bigl(\DiK^{(0)}_{(+)}+\DiK^{(0)}_{(-)}\Bigr)=2-0\;,
\end{equation}
which equals the Euler number of the sphere. 
The index could also be obtained as the difference between the even
and the odd zero modes of $\DiK$.\\
The spectrum of the Dirac operator is obtained solving the eigenvalue
equation :
\begin{eqnarray}
\j\,\Bigl(
f_A\,{\partial\over\partial\zeta_A^c}\,-\,\zeta_A\Bigr)
\psi^{A,(1)}_+&=&\lambda \psi^{A,(0)}_+\;\nonumber\\
\j\,\Bigl(f_A\,{\partial\over\partial\zeta_A}\Bigr) 
\psi^{A,(0)}_+&=&\lambda \psi^{A,(1)}_+\;,\label{eigeq1}
\end{eqnarray}
which yields
\begin{equation}
\label{eigeq2}
-f_A^2\,{\partial^2\over \partial\zeta_A\partial\zeta_A^c}
\psi^{A,(0)}_+=\lambda^2\psi^{A,(0)}_+\;.
\end{equation}
To make a link with wellknown facts, it is useful to introduce the Killing
vector fields generating the $SO(3)$ action on $S^2$ :
\begin{eqnarray}
\widehat{\ell}_+ & = & 
-\Bigl({\partial\over\partial\zeta_A^c}+
\zeta_A^2{\partial\over\partial\zeta_A}\Bigr)\;,\nonumber\\
\widehat{\ell}_- & = &
+\Bigl({\partial\over\partial\zeta_A}+
(\zeta_A^c)^2{\partial\over\partial\zeta_A^c}\Bigr)\;,\nonumber\\
\widehat{\ell}_z & = &
+\Bigl(\zeta_A
{\partial\over\partial\zeta_A}-
\zeta_A^c{\partial\over\partial\zeta_A^c}\Bigr)\;,\label{angmom}
\end{eqnarray}
with the Casimir 
\begin{equation}
\label{casimir}
\widehat{L}^2= 
{1\over 2}\Bigl(\widehat{\ell}_+\widehat{\ell}_-+\widehat{\ell}_-\widehat{\ell}+\Bigr)
+{\widehat{\ell}_z}^2=
-f_A^2\;{\partial^2\over\partial\zeta_A\partial\zeta_A^c}\;.
\end{equation}   
It follows that the eigenvalues are $\lambda=\pm \sqrt{\ell(\ell+1)}$ 
with multiplicity $(2\ell + 1)$ and eigenvectors given by 
\begin{equation}
\label{eigenvec}
\psi^A_{(+)}(\pm,\ell,m)=
\left(
\begin{array}{c}
\psi^{A,(0)}_+(\pm,\ell,m)\\
\psi^{A,(1)}_+(\pm,\ell,m)
\end{array}
\right)\;,
\end{equation}
with
\begin{eqnarray*}
\psi^{A,(0)}_+(\pm,\ell,m) & = &{\rm Y}_{\ell,m}\;,\\
\psi^{A,(1)}_+(\pm,\ell,m) & = &  
{\pm\j\over\sqrt{\ell(\ell+1)}}
\Bigl(\hat\ell_-+\zeta_A^c\hat\ell_z\Bigr){\rm Y}_{\ell,m}\\
  & = & 
{\pm\j\over\sqrt{\ell(\ell+1)}}
\Bigl(\sqrt{(\ell+m)(\ell-m+1)}{\rm Y}_{\ell,m-1}
+m\,\zeta_A^c\,{\rm Y}_{\ell,m}\Bigr)
\end{eqnarray*}
\section{The Projective Modules over $\diff\otimes{\bf C}$}
\label{Pmod}
\setcounter{equation}{0}

Let $\Mod$ be a (right) module 
over the algebra\footnote{Troughout this section,
elements of the algebra $\Al$ are denoted by
$\bigl\{a,b,\cdots\bigr\}$, vectors of the module $\Mod$ by 
$\bigl\{X,Y,\cdots\bigr\}$ and complex numbers by 
$\bigl\{\kappa,\lambda,\cdots\bigr\}$}
$\Al=\diff\otimes{\bf C}$ with involution 
$ a\rightarrow a^+$
given here by complex conjugation $a^+=a^c$.\\
Assume $\Mod$ to be endowed with a sesquilinear form $\h$, i.e. a  
mapping 
\begin{equation}
\label{modhermit}
\h :\Mod\times\Mod\rightarrow \Al:X,Y\rightarrow \h(X,Y)\;,
\end{equation}
which is bi-additive  and obeys $\h(Xa,Yb)= a^+\,\h(X,Y)\,b\;$.\\
It is nondegenerate if $\h(X,Y)=0,\forall X\Rightarrow Y=0$ and\\
$\h(X,Y)=0,\forall Y\Rightarrow X=0$. It is hermitian if 
$\h(X,Y)=\Bigl(\h(Y,X)\Bigr)^+$ 
and positive definite if it is hermitian 
and if $\h(X,X)$ is a positive element of $\Al$ for all $X\neq 0$.\\
The adjoint,  ${\bf S}^\dag$, with respect to $\h$ of an endomorphism 
${\bf S}\in\hbox{END}_\Al(\Mod)$ is defined by 
$\h(X,{\bf S}^\dag Y)=\h({\bf S}X,Y)\,$.\\
The universal graded differential envelope of $\Al$ 
(see e.g.\cite{Coq2},\cite{V-GB})
is the graded differential algebra
\begin{equation}
\label{Omega}
\Bigl\{\Uda^\bullet(\Al)=\bigoplus_{k=0}^\infty\,
\Uda^{(k)}(\Al)\,,\,\du\Bigr\}\;.
\end{equation}
Its most pragmatic definition goes as follows.
Let $\du\Al$ be a copy of $\Al$ as a set and consider 
the free algebra over $\bf C$ generated
by the elements of $\Al$ and $\du\Al$, then 
$\Uda^\bullet(\Al)$ can be defined as this free algebra modulo 
the relations 
\begin{eqnarray*}
\kappa \du a + \lambda \du b & =& \du(\kappa a + \lambda b)\;,\\
\du(a\,b)& =& (\du a)\,b + a\,(\du b)\;,\\
\du 1& =& 0\;.
\end{eqnarray*}
The involution in $\Al$ can be extended to $\Uda^\bullet(\Al)$ such 
that\footnote{Several authors, e.g. Connes in \cite{Con1}, 
use a different convention :
$\Bigl(\du a\Bigr)^+=-\,\du a^+\;$.}
\begin{equation}
\label{invol}
\Bigl(\du a\Bigr)^+=\du a^+\;,
\end{equation}
and, with $\ua_u$ be the main automorphism (see (\ref{mainauto})) of 
the graded algebra $\Uda^\bullet(\Al)$ ,  one obtains 
\begin{equation}
\label{unimainauto}
\forall \psi_u\in\Uda^\bullet(\Al)\;,\;\Bigl(\du\psi_u\Bigr)^+=-\ua_u\;\du\Bigl(\psi_u^+\Bigr)\;.
\end{equation}
In fact, $\Uda^\bullet(\Al)$ is an $\Al$-bimodule so that its tensor product
over $\Al$  with  $\Mod$, $\Omega_u^\bullet(\Mod)=\Mod\,\otimes_a\,\Uda^\bullet(\Al)$, 
is well defined. 
The hermitian structure $\h$ of  (\ref{modhermit}) can be extended to  $\Omega_u^\bullet(\Mod)$ by :
\begin{eqnarray}
\h& : &\Omega_u^\bullet(\Mod)\times\Omega_u^\bullet(\Mod)
\rightarrow
\Uda^\bullet(\Al):\nonumber\\
& & \Bigl(X\otimes_a\psi_u\,,\,Y\otimes_a\phi_u\Bigr) \rightarrow
\h(X\otimes_a\psi_u\,,\,Y\otimes_a\phi_u)=
\psi_u^+\,\h(X,Y)\,\phi_u\;.\label{unihermit}
\end{eqnarray}
A connection in $\Mod$ is a maping
$\nabla:\Mod\rightarrow\Omega_u^{(1)}(\Mod):X\rightarrow\nabla X\;,$
such that :
\begin{eqnarray}
\nabla\bigl(X+Y\bigr)& = & \nabla X + \nabla Y \;,\nonumber\\
\nabla(X\,a)& = & (\nabla X)a+X\otimes_a\du a\;.\label{modconnection}
\end{eqnarray}
The connection also can be extended to $\Omega_u^\bullet(\Mod)$ :
\begin{eqnarray}
\nabla& : & \Omega_u^{(k)}(\Mod)\rightarrow\Omega_u^{(k+1)}(\Mod):\nonumber\\
&  & X\otimes_a\psi_u\rightarrow
\nabla(X\otimes_a\psi_u)=(\nabla X)\psi_u+X\otimes_a\du\psi_u\;.
\label{uniconnection}
\end{eqnarray}
The square of the connection is its curvature which is a 
right module homomorphism :
\begin{equation}
\label{curvature}
\nabla^2 : \Omega_u^{(k)}(\Mod)\rightarrow\Omega_u^{(k+2)}(\Mod)
\;\mbox{, i.e.}\;
\nabla^2(X\otimes_a\psi_u\,a) = \Bigl(\nabla^2(X\otimes_a\psi_u)\Bigr)\,a\,.
\end{equation}
The connection is compatible with the hermitian structure if
\begin{equation}
\label{hermitconnection}
\du\h(X,Y)=\h(\nabla X,Y)+\h(X,\nabla Y)\;.
\end{equation}
This is easily extended to $\Omega_u^\bullet(\Mod)$:
$$\forall \widehat{\psi_u},\,\widehat{\phi_u}\in \Omega_u^\bullet(\Mod),\;
\du\h(\ua_u(\widehat{\psi_u}),\widehat{\phi_u})
=\h(\nabla \widehat{\psi_u},\widehat{\phi_u})
+\h(\widehat{\psi_u},\nabla\widehat{\phi_u})\;.$$
The curvature is then always a hermitian operator :
\begin{equation}
\label{hermitcurvature}
\h(\nabla^2 X,Y)-\h(X,\nabla^2 Y)=0\,.
\end{equation}
A free module of finite rank is a module isomorphic to $\Al^N$ and has a basis
$\bigl\{E_i\,;\,i=1,\cdots,N\bigr\}$
so that each element of $\Mod$ can be written as $X=E_i\,x^i$.
The hermitian structure is then given by
$\h(X,Y)=(x^i)^+\,h_{\bar i\,j}\,y^j\;.$\\
By definition, in a standard unitary basis :
$h_{\bar i\,j}=\left\{
\begin{array}{ll}
1 & \mbox{if i= j ,}\\
0 & \mbox{otherwise}.
\end{array}
\right.\;$\\
In the basis $\bigl\{E_i\bigr\}$, the connection $\nabla$ is given by 
a $N\times N$ matrix with entries in $\Uda^{(1)}(\Al)$ :
\begin{equation}
\label{freeconnection}
\nabla E_i=E_j\otimes_a{\omega^j}_i\;,
\end{equation}
where ${\omega^j}_i\,\in\Uda^{(1)}(\Al)\;,$ so that 
\begin{equation}
\label{covariantder}
\nabla X=E_i\otimes_a\,(\du x^i+{\omega^i}_j\,x^j)\;.
\end{equation}
The curvature of the connection is then given by : 
\begin{equation}
\label{freecurvature}
\nabla^2\,E_i=E_j\otimes_a\,{\rho^j}_i\;;\;
{\rho^j}_i=\du{\omega^j}_i+{\omega^j}_k\,{\omega^k}_i\in
\Uda^{(2)}(\Al)\;.
\end{equation}
The compatibility with the hermitian structure reads: 
\begin{equation}
\label{hermitfreeconnection}
\du h_{\bar i j}=\Bigl({\omega^k}_i\Bigr)^+\,h_{\bar k j}+
h_{\bar i \ell}\,{\omega^\ell}_j\;,
\end{equation}
\begin{equation}
\label{hermitfreecurvature}
\Bigl({\rho^k}_i\Bigr)^+\,h_{\bar k j}-
h_{\bar i \ell}\,\Bigl({\rho^\ell}_i\Bigr)=0\;.
\end{equation}
An endomorphism ${\bf S}\in\hbox{END}_\Al(\Mod)$, given by
${\bf S}E_i=E_j\,s^j_i$, has adjoint ${\bf S}^\dagger$  given by 
${\bf S}^\dagger E_i=E_j\,(s^\dagger)^j_i$  such that
$h_{\bar i\,k}\,(s^{\dagger})^ k_j=({s^\ell}_i)^+\,h_{\bar \ell\,j}$.\\
A hermitian projective module of finite rank $\Mod_{\Proj}$ over $\Al$ 
is obtained from a free module $\Mod$ as the image 
of a hermitian projection operator $\Proj\in\hbox{END}_\Al(\Mod)$ such that 
$\Proj^2=\Proj$ and $\Proj^\dagger=\Proj$.\\
An element $X\in\Mod$ belongs to $\Mod_\Proj= \Proj\,\Mod$  iff
$\Proj X = X$.\\
The hermitian structure $\h$ in $\Mod$ defines
an hermitian structure $\h_\Proj$ in $\Mod_\Proj$ by restricting 
$X,Y$ to $\Mod_\Proj$ :
\begin{equation}
\label{projhermit}
\h_\Proj(X,Y)=\h(X,Y)=\h(\Proj X,Y)=\h(X,\Proj Y)=
\h(\Proj X,\Proj Y)\;.
\end{equation}
Such a projection operator can be expanded in terms of the identity
and the Pauli--Gell-Mann hermitian traceless matrices 
$\bigl\{\lambda_\alpha\;;\;\alpha=1,\cdots,N^2-1\bigr\}$ as 
\begin{equation}
\label{generalprojector}
\Proj={1\over 2}\bigl(a\,\um\,+\,b^\alpha\,\lambda_\alpha\bigr)\;,\;
\end{equation}
where $a\;\hbox{and the}\;b^\alpha\;$ are real valued functions on the 
manifold (here $S^2$).\\
Using the multiplication of the $\lambda$ matrices in standard notation :\\
$\lambda_\alpha\lambda_\beta=
{2\over N}\delta_{\alpha\beta}\,+\,
\lambda_\gamma
\bigl(^\gamma d_{\alpha\beta}\,+\,^\gamma f_{\alpha\beta}\bigr)\;,$
leads to :
\begin{eqnarray*}
a^2+{2\over N}\delta_{\alpha\beta}\,b^\alpha b^\beta 
&=& 2\,a\\
2\,a\,b^\alpha\,+\,^\gamma d_{\alpha\beta}\,b^\alpha b^\beta 
&=& 2\,b^\gamma
\end{eqnarray*}
It can be shown that, in the case of the sphere, it is enough to consider only $N=2$ so that, besides
the trivial solutions $\Proj=\um$ and $\Proj=0$, the general 
solution is obtained as :
\begin{equation}
\label{sphereprojector}
\Proj(\vec n)={1\over 2}\bigl(\um+n^\alpha\sigma_\alpha\bigr)\;,
\end{equation}
where $n^\alpha$ are real functions on the sphere such that
$\delta_{\alpha\beta}\,n^\alpha\,n^\beta\,=\,1$.\\
The projection operators are thus given by mappings 
$ S^2\rightarrow S^2$ and it can be shown
that homotopic mappings define isomorphic projective modules.
These are thus classified by the second homotopy
group $\pi_2(S^2)\simeq{\bf Z}$.\\
In each homotopy class $\bigl[\vec n\bigr]$, a representative 
can be choosen in various ways.  Our choice is the following.
First we choose coordinates in the domain sphere and in the 
target sphere such that  the point (1,0,0) of the domain is mapped 
on the fixed base point in the target sphere and, in this target sphere,
coordinates are chosen such that the fixed base point is also
given by (1,0,0).\\
Let $H^{\,im}_A$ and $H^{\,im}_B$ be the 
austral and boreal charts of the image sphere with complex coordinates
$\nu_A\,,\,\nu_B$, then
in each homotopy class $\bigl[\vec n\bigr]$ we choose :
\begin{eqnarray}
\nu_A=\bigl(\zeta_A\bigr)^\n\;,&
\;\nu_B=\bigl(\zeta_B\bigr)^\n\;,&
\hbox{if}\;\bigl[\vec n\bigr]= +\n\,;\label{positivehomotopy}\\
\nu_A=\bigl(\zeta_A^c\bigr)^\n\;,&
\;\nu_B=\bigl(\zeta_B^c\bigr)^\n\;,& 
\hbox{if}\;\bigl[\vec n\bigr]=-\n\;.\label{negativehomotopy}
\end{eqnarray}
In this way, $H^{\,im}_A$ and $H^{\,im}_B$ are the images of
$H_A$ and $H_B$.\\
In a standard unitary basis $\bigl\{E_i\,;\,i=1,2\bigr\}$ of 
$\Al^2$ the projection operator 
$\Proj E_i=E_j\,{p^j}_i$, 
with the usual representation of the Pauli matrices, 
is given by the matrix  :
\begin{equation}
\label{basissphereprojector}
\MProj={1\over 2}
\left(
\begin{array}{cc}
1+n_z & n_x-\j n_y \\
n_x+\j n_y & 1-n_z
\end{array}
\right)\;.
\end{equation}
It is given in $H^{\,im}_A$, respectively in $H^{\,im}_B$, by\footnote
{Here we obtain, in a quite natural way, the so-called Bott  projection, 
used in algebraic K-theory \cite{Wegge}. 
We thank J.M. Gracia-Bond\'{\i}a for pointing this out.}:
\begin{equation}
\label{Bottprojector}
\MProj_A={1\over 1+\vert\nu_A\vert^2}\,
\left(
\begin{array}{cc}
\vert\nu_A\vert^2 & \nu_A^c \\
\nu_A & 1
\end{array}
\right)\;;\;
\MProj_B=
{1\over 1+\vert\nu_B\vert^2}\,
\left(
\begin{array}{cc}
1 & \nu_B\\
\nu_B^c & \vert\nu_B\vert^2 
\end{array}
\right)\;.
\end{equation} 
An element $X=E_i\,x^i\in \Al^2$, given by the 
column matrix ${\cal X}=\left(\begin{array}{c}
x^1\\
x^2
\end{array}\right)$,
belongs to $\Mod_\Proj$ if
$\MProj {\cal X}={\cal X}$. Here $x^1,x^2$ are functions 
on $S^2$, given in the chart $H_A$ by 
functions $x^1_A,x^2_A$ of $(\zeta_A,\zeta_A^c)$, related by
$x^1_A=\nu_A^c\,x^2_A\;.$
In $H_B$ they are given by $x^1_B,x^2_B$, with 
the relation 
$x^2_B=\nu_B^c\,x^1_B\;.$\\
A local, normalized basis is defined  in $H_A$, by :
\begin{equation}
\label{localbasisa}
E_A=
\left(\begin{array}{cc}E_1 & E_2\end{array}\right)\;{\cal E}_A\;;\;
{\cal E}_A=
\left(\begin{array}{c}\nu_A^c \\ 1\end{array}\right)\;
{1\over\sqrt{1+\vert\nu_A\vert^2}}\;,
\end{equation}
and, in $H_B$ by :
\begin{equation}
\label{localbasisb}
 E_B=
\left(\begin{array}{cc}E_1 & E_2\end{array}\right)\;{\cal E}_B\;;\;
{\cal E}_B=
\left(\begin{array}{c}1 \\ \nu_B^c\end{array}\right)\;
{1\over\sqrt{1+\vert\nu_B\vert^2}}\;.
\end{equation} 
In $H_A\cap H_B$, they are related by the (passive) gauge transformation :
\begin{equation}
\label{gaugetransform4}
E_A=
E_B\;g^B_A\quad,\;\hbox{ with}\;\;g^B_A={\nu_A^c\over\vert\nu_A\vert}\;.
\end{equation}
In these local bases an element ${\cal X}\in \Mod_\Proj$ is written as
${\cal X}={\cal E}_A\,x^A$ in $H_A$ and 
${\cal X}={\cal E}_B\,x^B$ in $H_B$ with :
\begin{equation}
\label{gaugetransform5}
x^B= g^B_A\,x^A\;.
\end{equation} 
The hermitian product of two elements $X$ and $Y$ of $\Mod_\Proj$
reads 
\begin{eqnarray}
h_\Proj(X,Y)& =&(x^A)^+\,y^A\;\;\hbox{in}\;H_A\;\nonumber\\
                     &=& (x^B)^+\,y^B\;\;\hbox{in}\;H_B\;.\label{localprojhermit}
\end{eqnarray}
A connection $\nabla$ in $\Mod$ induces a connection $\nabla_\Proj$ 
in $\Mod_\Proj$.
\begin{equation}
\label{projconnection}
\nabla_\Proj = \Proj\,\circ\,\nabla :
\Mod_\Proj\rightarrow\Mod_\Proj\otimes_a\Uda^{(1)}(\Al)\;,
\end{equation}
such that
\begin{equation}
\label{projcovariantderiv1}
\nabla_\Proj X=E_i\otimes_a\,\Bigl({p^i}_j\,\du x^j + (p^i_k\,\omega^k_\ell\,p^\ell_j)\; x^j\Bigr)\;,
\end{equation}
In matrix notation, this reads :
\begin{equation}
\label{projcovariantderiv2}
\nabla_\Proj {\cal X}=
\Bigl(\MProj \du {\cal X}+\Bigl(\MProj (\omega)\MProj\Bigr){\cal X}\Bigr)\;.
\end{equation}
With $(\kappa)=\Bigl(\MProj (\omega)\MProj\Bigr)$, the curvature ${\nabla_\Proj}^2$ reads:
\begin{equation}
\label{projcurvature}
{\nabla_\Proj}^2 {\cal X}=
\Bigl(\MProj (\du(\kappa))\MProj+(\kappa)^2+
\MProj(\du\MProj)(\du\MProj)\MProj\Bigr) {\cal X}\;.
\end{equation}   
A general hermitian connection in $\Al^2$ is given by :
$$(\omega)={1\over\j}\,
\left(\begin{array}{c c}
\omega_1 & \sigma\\
\sigma^+ & \omega_2\end{array}\right)\;,$$
with $\omega_1,\omega_2$ and $\sigma$ in $\Uda^{(1)}(\Al)$ and
${\omega_1}^+=\omega_1\;,\;{\omega_2}^+=\omega_2\;$.\\
The matrix $(\kappa)$ is given in $H_A$ by :
\begin{equation}
\label{kappainA}
(\kappa)=
{1\over\sqrt{1+\vert\nu_A\vert^2}}\,
\left(
\begin{array}{c c}
\nu_A^c\,\kappa_A\,\nu_A & \nu_A^c\,\kappa_A\\
\kappa_A\,\nu_A      & \kappa_A
\end{array}
\right)\,
{1\over\sqrt{1+\vert\nu_A\vert^2}}\;,
\end{equation}
where
$$\kappa_A={1\over\sqrt{1+\vert\nu_A\vert^2}}\,
{1\over\j}
\Bigl(
\nu_A\,\omega_1\,\nu_A^c+\nu_A\,\sigma+\sigma^+\,\nu_A^c+\omega_2
\Bigr)\,
{1\over\sqrt{1+\vert\nu_A\vert^2}}\;;$$
and in $H_B$ by :
\begin{equation}
\label{kappainB}
(\kappa)=
{1\over\sqrt{1+\vert\nu_B\vert^2}}\,
\left(
\begin{array}{c c}
\kappa_B           & \kappa_B\,\nu_B\\
\nu_B\,\kappa_B  & \nu_B^c\,\kappa_B\,\nu_B
\end{array}
\right)\,
{1\over\sqrt{1+\vert\nu_B\vert^2}}\;,
\end{equation}
where
$$\kappa_B={1\over\sqrt{1+\vert\nu_B\vert^2}}\,
{1\over\j}\Bigl(
\omega_1+\nu_B\,\sigma^++\sigma\,\nu_B^c+\nu_B\,\omega_2\,\nu_B^c
\Bigr)\,
{1\over\sqrt{1+\vert\nu_B\vert^2}}\;.$$
In $H_A$, on ${\cal X}={\cal E}_A\,x^A$, one obtains :
\begin{equation}
\label{localcovarderivA}
\nabla_\Proj {\cal X}_{\,\vert A}
={\cal E}_A\,\Bigl(\du x^A +\gamma_A x^A\Bigr)\;,
\end{equation}
with the (universal) total potential
$\gamma_A =\kappa_A+\mu_A\;$, where
\begin{equation}
\label{monopoleA}
\mu_A=
{1\over\sqrt{1+\vert\nu_A\vert^2}}
\Bigl(\nu_A\,\du\nu_A^c-
\sqrt{1+\vert\nu_A\vert^2}\,
\du\sqrt{1+\vert\nu_A\vert^2}\Bigr)
{1\over\sqrt{1+\vert\nu_A\vert^2}}
\end{equation}
will be called the (universal) monopole potential.\\
In the same way, in $H_B$,
on ${\cal X}={\cal E}_B\,x^B$, one obtains :
\begin{equation}
\label{localcovarderivB}
\nabla_\Proj {\cal X}_{\,\vert B}
={\cal E}_B\,\Bigl(\du x^B +\gamma_B x^B\Bigr)\;,
\end{equation}
with $\gamma_B =\kappa_B+\mu_B\;$, where
\begin{equation}
\label{monopoleB}
\mu_B=
{1\over\sqrt{1+\vert\nu_B\vert^2}}
\Bigl(\nu_B\,\du\nu_B^c-
\sqrt{1+\vert\nu_B\vert^2}\,
\du\sqrt{1+\vert\nu_B\vert^2}\Bigr)
{1\over\sqrt{1+\vert\nu_B\vert^2}}\;.
\end{equation}
The curvature operator reads 
\begin{eqnarray}
{\nabla_\Proj}^2{\cal X}_{\,\vert A}&=&
{\cal E}_A\,{\cal R}_A\,x^A\;,\;\hbox{with}\;
{\cal R}_A=\du\,\gamma_A+\gamma_A\gamma_A\;,\nonumber\\
{\nabla_\Proj}^2{\cal X}_{\,\vert B}&=&
{\cal E}_B\,{\cal R}_B\,x^B\;,\;\hbox{with}\;
{\cal R}_B=\du\,\gamma_B+\gamma_B\gamma_B\;.\label{localcurvature}
\end{eqnarray}
The gauge transformations in the overlap region 
$H_A\cap H_B$ are :
\begin{eqnarray}
\kappa_A&=&\Bigl(g^B_A\Bigr)^{-1}
\kappa_B\; g^B_A\;,\nonumber\\
\mu_A&=&\Bigl(g^B_A\Bigr)^{-1}
\mu_B\;g^B_A
+
\Bigl(g^B_A\Bigr)^{-1}
\du g^B_A\,,\nonumber\\
\gamma_A&=&\Bigl(g^B_A\Bigr)^{-1}
\gamma_B\;g^B_A
+
\Bigl(g^B_A\Bigr)^{-1}
\du g^B_A\;,\label{gaugetransform6}\\
{\cal R}_A&=&\Bigl(g^B_A\Bigr)^{-1}
{\cal R}_B\;g^B_A\;.\label{gaugetransform7}
\end{eqnarray}
%
\section{ The spectral triple $\Bigl\{\Al,\Hi,\DiK\Bigr\}$}
\label{triple}
\setcounter{equation}{0}

The spectral triple $\Bigl\{\Al,\Hi,\DiK\Bigr\}$, as defined by 
Connes \cite{Con2,Con3}, is given here by the algebra 
$\,\Al=\diff\otimes{\bf C}\;$ 
with complex conjugation as involution and 
$\Vert f\Vert_\Al = {\bf sup}_{x\in S^2}\vert f(x)\vert\;$ 
as  ${\bf C}^*$-algebra norm.
The Hilbert space $\Hi$ is given by the completion $\Hi(P_+)$ of the left ideal  
$\Id^E_+$ of sections in the Clifford algebra bundle with inner product :
\begin{equation}
\langle \Psi,\Phi\rangle = \int_{S^2}\,\Psi^c\wedge\star\Phi
\end{equation}
On this Hilbert space, there is a faithful $\ast$-representation $\pi_0$ 
of $\Al$ given here by pointwise multiplication:
\begin{eqnarray}
\pi_0&:\Al\rightarrow\L(\Hi)&:f\rightarrow\widehat{f} =\pi_0(f)\nonumber\\
 & \Bigl(\widehat{f}\Psi)(x)&= f(x)\,\Psi(x)
\end{eqnarray}
Faithfulness implies that operator norm and ${\bf C}^*$-algebra norm coincide :
$\Vert \hat f\Vert = \Vert f\Vert_\Al\;.$\\
The unbounded, essentially selfadjoint, Dirac operator is  
$\DiK=\j(\dd-\delta)$,  restricted to $\Hi(P_+)$.
Its spectrum is
$\Bigl\{\pm\sqrt{\ell(\ell+1)}\;;\;\ell=0,1,2,\cdots\Bigr\}\,,$
so that the resolvent $(\DiK-z)^{-1}\;,\;z\not\in{\bf R}\,,$ is compact.\\
Furthermore, its commutator with $\hat f$ is calculated 
using formula (\ref{difcodif}) :
\begin{eqnarray*}
\Bigl[{1\over \j}\DiK,\hat f\Bigr]\Psi
&=&(\dd -\delta) f\Psi- f(\dd-\delta)\Psi\\
&=&
\ut^\mu\vee\bigl(\nabla_{\vec e_\mu}f\Psi\bigr)
- f\ut^\mu\vee\bigl(\nabla_{\vec e_\mu}\Psi\bigr)\\
&=&
\ut^\mu\vee\bigl(\vec e_\mu(f)\bigr)\Psi\bigr)
=\dd f\vee\Psi
\end{eqnarray*}
It is bounded, with norm squared:
\begin{equation}
\Vert\Bigl[{1\over \j}\DiK,\hat f\Bigr]\Vert^2= 
{\bf sup}_{x\in S^2}\;h^{-1}(\dd f,\dd f)\;.
\end{equation}
The eigenvalues of $\vert\DiK\vert$ are 
$\Bigl\{\sqrt{\ell(\ell+1)}\;;\;\ell=0,1,2,\cdots\Bigr\}\,,$
with multiplicities 
\begin{equation}
\mu_\ell=\left\{
\begin{array}{ll}
1 & \mbox{if $\ell=0$}\\
2(2\ell+1) & \mbox {if $\ell\neq 0$}
\end{array}
\right.
\nonumber
\end{equation}
These eigenvalues are ordered in an increasing sequence so that 
the order number $n_L$ of the eigenvalue $\lambda_{n_L}=\sqrt{L(L+1)}$,
counting the multiplicity, is
$$n_L=1+ \sum_{\ell=1}^{L-1}\;2(2\ell+1)= 2L^2-1\;.$$
It follows then that the order of 
$\lambda_{n_L}$ as $n_L\rightarrow\infty$ is $(n_L)^{1/2}$. In Connes' terminology, 
the spectral triple is called $(d,\infty)$ summable with $d=2$.
It can then be shown that the Dixmier trace\footnote{ We refer to \cite{Con1,V-GB} for 
the definition and properties of the Dixmier trace.} 
${\bf Tr}_{Dix}\Bigl(\hat f \vert \DiK\vert^{-d}\Bigr)$
exists and, in our case, is given by :
\begin{equation}
\label{Dix1}
{\bf Tr}_{Dix}\Bigl(\hat f \vert D\vert^{-2}\Bigr)= {1\over 2\pi}\int_{S^2}\,f\,\omega
\end{equation}
The Dirac operator extends $\pi_0$ to a $\ast$-representation $\pi$ of 
$\Uda^\bullet(\Al)$ in $\Hi$:
\begin{eqnarray}
\pi:&\Uda^\bullet(\Al)\rightarrow\L(\Hi)&:
\psi_u\rightarrow \pi(\psi_u)\;,\nonumber\\
&\pi(f_0\,\du f_1\,\cdots\,\du f_k)&=
\widehat{ f_0}\,\Bigl[{1\over \j}\DiK,\widehat{f_1}\Bigr]\,\cdots\,
\Bigl[{1\over \j}\DiK,\widehat{f_k}\Bigr]\nonumber\\
\hbox{(given here by )}&&=
f_0\,\dd f_1\,\vee\,\cdots\,\vee\,\dd f_k\,\vee\;.\label{piomega}
\end{eqnarray}
However, this representation is not differential since 
$\psi_u\in\hbox{Ker}(\pi)$ does \underbar{not} imply 
$\pi(\du \psi_u)=0$, which is easily seen by the standard example:
$$\pi\Bigl(2f\du f-\du(f^2)\Bigr)=\Bigl(2f\dd f-\dd(f^2)\Bigr)=0\;,\;
\hbox{while}$$
$$\pi\Bigl(\du(2f\du f-\du(f^2))\Bigr)=\pi\Bigl(2\du f\,\du f\Bigr)
=2\dd f \vee \dd f= 2 g^{-1}(\dd f,\dd f)\,{\bf Id}\,.$$
To obtain a graded differential algebra of operators 
in $\Hi$ one has to take the quotient of $\Uda^\bullet(\Al)$ 
by the graded differential ideal, often called "junk",
\begin{equation}
\label{defjunk}
{\cal J}={\cal J}_0 + \du\Bigl({\cal J}_0\Bigr)=
\bigoplus_{k=0}^{\infty}\Bigl({{\cal J}_0}^{(k)}+\du{{\cal J}_0}^{(k-1)}\Bigr)\;, 
\end{equation}
where ${\cal J}_0=\hbox{Ker}(\pi)$.
In this way, one obtains the graded differential algebra :
\begin{equation}
\label{nojunk}
\UDda^\bullet(\Al)={\Uda^\bullet(\Al)\over{\cal J}}
=\bigoplus_{k=0}^{\infty}\UDda^{(k)}(\Al)\;,
\end{equation} 
with canonical projection
\begin{equation}
\label{junkproj}
\pi_D:
\Uda^\bullet(\Al)\rightarrow
\UDda^\bullet(\Al)
\end{equation}
The classical homomorphism theorem, applied to the representation $\pi$, yields the 
isomorphism
\begin{equation}
\label{omegaD}
\UDda^{(k)}(\Al)\cong 
{\pi\Bigl(\Uda^{(k)}(\Al)\Bigr)\over
\pi\Bigl(\du{{\cal J}_0}^{(k-1)}\Bigr)}\;.
\end{equation}
In $\pi\Bigl(\Uda^{(k)}(\Al)\Bigr)$, the scalar product 
of $R$ and $S$, belonging to $\pi\Bigl(\Uda^{(k)}(\Al)\Bigr)$, 
is defined by 
\begin{equation}
\label{scalarproduniv }
\langle R,S \rangle_{(k)}={\bf Tr}_{Dix}\Bigl(R^\dagger\,S\vert \DiK\vert^{-d}\Bigr)\;.
\end{equation}
In the corresponding Hilbert space completion 
$\Hi^{(k)}_u$ of $\pi\Bigl(\Uda^{(k)}(\Al)\Bigr)$, let
$\MProj^{(k)}$ be the projector on the orthogonal complement
of $\pi(\du{{\cal J}_0}^{(k-1)})$.\\
Then, $\langle\MProj^{(k)}R,\MProj^{(k)}S\rangle_{(k)}$ depends 
only on the equivalence classes of $R$ and $S$ in $\UDda^{(k)}(\Al)$ 
and defines a scalar product in 
$\Hi^{(k)}_D=\MProj^{(k)}\Hi^{(k)}_u=
\Bigl(\pi(\du{{\cal J}_0}^{(k-1)})\Bigr)^{\perp}$  which contains 
$\UDda^{(k)}(\Al)$ as a dense subspace.
Indeed, let $R_D$ and $S_D$ belong to $\UDda^{(k)}(\Al)$, which means that
$R_D$ and $S_D$ are equivalence classes of elements $R$ and $S$ in
$\Uda^{(k)}(\Al)$ modulo $\pi(\du{{\cal J}_0}^{(k-1)})$, then 
\begin{equation}
\label{omegadproduct} 
\langle R_D,S_D\rangle_{(k),D}=\langle\MProj^{(k)}R,\MProj^{(k)}S\rangle_{(k)}\;.
\end{equation}
Furthermore, with this scalar product, each $\Hi^{(k)}_D$ is endowed with a left- and 
a right representation of the unitary group $u^+u=u u^+=\um$ of the algebra $\Al$ :
\begin{equation}
\label{unitaryrep}
\langle\widehat{u}R_D,\widehat{u} S_D\rangle _{(k),D}=
\langle R_D,S_D\rangle _{(k),D}=\langle R_D\widehat{u},S_D\widehat{u}\rangle _{(k),D}\;.
\end{equation}
This follows from
$\langle\hat u R,\hat u S\rangle _{(k)}=
\langle R,S\rangle _{(k)}=
\langle R\hat u,S\hat u\rangle _{(k)}$, consequence of 
the asumed "tameness" (\cite{V-GB}) of the Dixmier trace :
\begin{equation}
\label{tameness}
{\bf Tr}_{Dix}\Bigl(\hat f R^\dagger \vert \DiK\vert^{-d}\Bigr)=
{\bf Tr}_{Dix}\Bigl(R^\dagger \hat f\vert \DiK\vert^{-d}\Bigr)\;;\;f\in \Al\;,
\end{equation}
and from the fact that $\MProj^{(k)}$ is a bimodule homomorphism :
\begin{equation}
\label{bimodhomo}
\MProj^{(k)}\Bigl(\widehat{f_1}R\widehat{f_2}\Bigr)=
\widehat{f_1}\Bigl(\MProj^{(k)}(R)\Bigr)\widehat{f_2}\;.
\end{equation}
It was shown by Connes (\cite{Con1,V-GB}) that, in the
commutative Riemannian case, this quotient 
amounts to identify
$\UDda^\bullet(\Al)$ with the usual de Rham algebra of differential forms.
This can be seen from formulae (\ref{Clifprod1}),(\ref{Clifprod2}), which yield
\begin{eqnarray*}
f_0\,\dd f_1\vee\dd f_2 &=&
f_0\,\dd f_1\wedge\dd f_2+ f_0\,g^{-1}(\dd f_1,\dd f_2)\\
&=&\;f_0\,\dd f_1\wedge\dd f_2+ {f_0\over 2}\Bigl(g^{-1}(\dd(f_1+f_2),\dd(f_1+f_2))\\
&& \;-g^{-1}(\dd f_1,\dd f_1)-g^{-1}(\dd f_2,\dd f_2)\Bigr).
\end{eqnarray*}
Now, $f\,g^{-1}(\dd h,\dd h)= f\,\dd h \vee \dd h = 
 \dd(f\,h)\vee\dd h + \dd({-f\over 2})\vee\dd(h^2)$ 
and, since $(f\,h)\dd h + ({-f\over 2})\dd(h^2)=0$, $f\,g^{-1}(\dd h,\dd h)$ belongs to 
$\pi\Bigl(\du{{\cal J}_0}^{(1)}\Bigr)$, so that 
\begin{equation}
\label{freejunk1}
\MProj^{(2)}\Bigl(f_0\,\dd f_1\vee\dd f_2\Bigr)= f_0\,\dd f_1\wedge\dd f_2\;.
\end{equation}
This is generalised to arbitrary $k$ :
\begin{equation}
\label{freejunkk}
\MProj^{(k)}\Bigl(f_0\,\dd f_1\vee\dd f_2\vee\cdots\vee\dd f_k\Bigr)=
f_0\,\dd f_1\wedge\dd f_2\wedge\cdots\wedge\dd f\;.
\end{equation}
Besides, the "Wick theorem" relating Clifford products with exterior products 
(see e.g. \cite{Kas}), resulting from (\ref{Clifprod1}) and (\ref{Clifprod2}), 
yields an explicit expression for the "junk" :
\begin{eqnarray}
\label{junk}
f_0\dd f_1\vee\dd f_2\vee&\cdots&\vee\dd f_k-
f_0\dd f_1\wedge\dd f_2\wedge\cdots\wedge\dd f_k
\nonumber\\&=&
\;+f_0\;g^{-1}(\dd f_1,\dd f_2)\,\dd f_3\wedge\dd f_4\wedge\cdots\wedge\dd f_k
\nonumber\\&& \; -
f_0\;g^{-1}(\dd f_1,\dd f_3)\,\dd f_2\wedge\dd f_4\wedge\cdots\wedge\dd f_k
\nonumber\\&& \; +\cdots\nonumber\\&&
\;+f_0\;g^{-1}(\dd f_1,\dd f_2)
\,g^{-1}(\dd f_3,\dd f_4)\,\dd f_5\wedge\cdots\wedge\dd f_k
\nonumber\\&& \; -\cdots\cdots\label{junkk}
\end{eqnarray}
The trace theorem (\cite{Con1,V-GB}) yields :
\begin{equation}
\label{tracedix}
\langle R_D,S_D\rangle_{(k),D}= 
{\bf Tr}_{Dix}\Bigl(\MProj^{(k)}R^\dagger\MProj^{(k)}S\vert\DiK\vert^{-2}\Bigl)=
{1\over 2\pi}\int_{S^2}\rho^c\wedge\ast\sigma\;,
\end{equation}
where $\rho$, $\sigma$ are the de Rham forms corresponding 
to $R_D$ and $S_D$.
\section{The Action}
\label{action}
To construct an action functional for the potential and the matter field, the 
projective modules $\Mod_\Proj=\Proj\Mod$ are tensored over $\Al$ with the Hilbert space $\Hi$, 
yielding a new Hilbert space :
\begin{equation}
\label{matter}
\Hi_F=\Mod_\Proj\otimes_a\Hi\;,
\end{equation}
whose generic elements, denoted by $\vert {\bf F}\rangle$, are linear combinations of 
the factorisable states $\vert X \otimes_a \Psi\rangle$, with $X\in\Mod_\Proj$ and
$\Psi\in \Hi$.\\
The scalar product, on these states, is defined by :
\begin{equation}
\label{scalarhf} 
\langle\!\langle X\otimes_a\Psi\parallel Y\otimes_a\Phi\rangle\!\rangle=
\langle\Psi,\pi_0\Bigl(\h_\Proj(X,Y)\Bigr)\Phi\rangle\;.
\end{equation}
The representation $\pi$ of $\Uda(\Al)$ (\ref{piomega})
induces a right $\Al$-module homomorphism
\begin{equation}
\label{righthomo}
{\cal R}_\pi:
\Uda^\bullet(\Mod_\Proj)=\Mod_\Proj\otimes_a\Uda^\bullet(\Al)
\rightarrow {\cal L}(\Hi,\Hi_F)\;,
\end{equation}
defined by :
$${\cal R}_\pi\Bigl(X\otimes_a\phi_u\Bigr)\Psi=
\vert X\otimes_a\pi(\phi_u)\Psi\rangle\;.$$
This homomorphism, in turn, induces a linear map
\begin{equation}
\label{linearmap}
{\cal O}_\pi:\hbox{HOM}_\Al\Bigl(\Mod_\Proj,\Uda^\bullet(\Mod_\Proj)\Bigl)
\rightarrow 
{\cal L}(\Hi_F): {\bf T}\rightarrow {\cal O}_\pi({\bf T})\;,
\end{equation}
defined by :
$${\cal O}_\pi({\bf T})\vert X\otimes_a \Psi\rangle=
\vert {\cal R}_\pi({\bf T}X)\Psi\rangle\;.$$
The Dirac operator in $\Hi$ and the connection $\nabla_\Proj$ in 
$\Mod_\Proj$ allow for the construction of a covariant Dirac operator 
$\DiK(\nabla_\Proj)$ in $\Hi_F$ :
\begin{equation}
\label{diraccovar}
\DiK(\nabla_\Proj)\vert X\otimes_a\Psi\rangle=
\vert X\otimes_a \DiK\Psi\rangle
+\j {\cal R}_\pi(\nabla_\Proj X)\Psi\;,
\end{equation}
which has a well defined action on 
$\vert X\,f\otimes_a\Psi\rangle=\vert X\otimes_a \widehat{f}\Psi\rangle$.\\
Furthermore, $\DiK(\nabla_\Proj)$ 
is formally self-adjoint in $\Hi_F$ and is 
covariant under the unitary automorphisms ${\bf U}$ of $\Mod_\Proj$ :
\begin{equation}
\label{covariance}
\DiK(^{\bf U}\nabla_\Proj){\bf U}\vert{\bf  F}\rangle=
{\bf U}\DiK(\nabla_\Proj)\vert {\bf F}\rangle\;,
\end{equation}
where
$^{\bf U}\nabla_\Proj= {\bf U}\circ\nabla_\Proj\circ{\bf U}^\dagger$.
\newpage\nin
The linear map (\ref{linearmap})
${\cal O}_\pi$ associates a selfadjoint operator ${\cal R}_F$  in $\Hi_F$ to
the curvature ${\nabla_\Proj}^2
\in \hbox{HOM}_\Al\Bigl(\Mod_\Proj,\Uda^{(2)}(\Mod_\Proj)\Bigl)$
so that the Dixmier trace \underline{in $\Hi_F$} defines a functional:
\begin{equation}
\label{yangmills1}
\Id\Bigl[\nabla_\Proj\Bigr]=
{\bf Tr}_{Dix}\Bigl({{\cal R}_F}^\dagger{\cal R}_F\;\vert \DiK(\nabla_\Proj)\vert^{-d}\Bigr)\;.
\end{equation}
On the other hand, the projection of $\nabla_\Proj$ with $Id\otimes \pi_D$ yields a connection 
$\nabla_\Proj^D$ on $\Mod_\Proj$ with values in  $\UDda^{(1)}(\Al)$.
In the basis $\Bigl\{E_i\Bigr\}$ of $\Mod$, this connection is given by (\ref{projconnection}):
\begin{equation}
\label{projcon}
\nabla_\Proj^D(\Proj E_i)= \Proj E_j\;{(\gamma)^j}_i\;,
\end{equation}
with  ${(\gamma)^i}_j\,\in\UDda^{(1)}(\Al)$, obeying ${(\gamma)^i}_j={\MProj^i}_k\,{(\gamma)^k}_\ell\,{\MProj^\ell}_j \;.$\\
The curvature $({\nabla_\Proj^D})^2$ is a homomorphism of 
$\hbox{HOM}_\Al\Bigl(\Mod_\Proj,\UDda^{(2)}(\Mod_\Proj)\Bigl)$, 
and is given by (\ref{projcurvature}):
\begin{equation}
\label{projcurv}
({\nabla_\Proj^D})^2\Proj\,E_i=\Proj E_j\,{{({\cal R}_D)}^j}_i\;,
\end{equation}
with\footnote{$\dd_D$ is the differential in 
$\UDda^\bullet(\Al)$.}
${{({\cal R}_D)}^j}_i=
{\MProj^i}_k\,\dd_D{(\gamma)^k}_\ell\,{\MProj^\ell}_j
+{(\gamma)^i}_k\,{(\gamma)^k}_\j
+{\MProj^i}_k\,\dd_D{\MProj^k}_\ell\,\dd_D{\MProj^\ell}_m\,{\MProj^m}_j\;.$\\
Let ${({\cal R})^j}_i$ be one of the operators  in $\Hi$ 
belonging to  $\pi\Bigl(\Uda^{(2)}(\Al)\Bigr)$, 
whose equivalence class is ${{({\cal R}_D)}^j}_i$. 
The Dixmier trace \underline{in $\Hi$} :
\begin{equation}
\label{yangmills2}
{\bf S}_{\bf YM}\Bigl[\nabla_\Proj^D\Bigr]=
\langle {{({\cal R}^\dagger_D)}^i}_j{{({\cal R}_D)}^j}_i \rangle_{(2),D}=
{\bf Tr}_{Dix}\Bigl(\MProj^{(2)}{({\cal R})^j}_i
\MProj^{(2)}{({\cal R})^i}_j \vert\DiK\vert^{-d}\Bigr)
\end{equation}
is well defined and, due to tameness (\ref{tameness}),  does not depend on the 
basis $\Bigl\{E_i\Bigr\}$ used in $\Mod$. According to Connes \cite{Con1}:
\begin{equation}
\label{yangmills3}
{\bf S}_{\bf YM}\Bigl[\nabla_\Proj^D\Bigr]=
\hbox{Inf}\Bigl\{
\Id\Bigl[\nabla_\Proj\Bigr]\;
;\;(Id\otimes \pi_D)\nabla_\Proj=\nabla_\Proj^D\Bigr\}
\end{equation}
The action for the matter field, living in $\Hi_F$, is given by :
\begin{equation}
\label{materie}
{\bf S}_{\bf F}\Bigl[\vert {\bf F}\rangle,\nabla_\Proj^D\Bigr]=
\langle{\bf F}\vert\DiK(\nabla_\Proj^D){\bf F}\rangle\;,
\end{equation}
where $\DiK(\nabla_\Proj^D)$ is the Dirac operator with connection $\nabla_\Proj^D$.
Finally, the total action for the matter field $\vert {\bf F}\rangle$ and the 
connection $\nabla_\Proj^D$,  is :
\begin{equation}
\label{actie}
{\bf S}\Bigl[\vert {\bf F}\rangle,\nabla_\Proj^D\Bigr]=
{\bf S}_{\bf F}\Bigl[\vert {\bf F}\rangle,\nabla_\Proj^D\Bigr]+
\lambda\;{\bf S}_{\bf YM}\Bigl[\nabla_\Proj^D\Bigr]\;,
\end{equation}
where $\lambda$ is a coupling constant.\\
This action and the resulting Euler-Lagrange equations will be written down
explicitely in our case study.\\
The connection $\nabla_\Proj^D$ in  
$\Mod_\Proj$ 
is obtained from (\ref{localcovarderivA}),(\ref{localcovarderivB}) as 
\begin{equation}
\label{covarderivd}
\nabla_\Proj^D{\cal E}_A={\cal E}_A\,\gamma_A^D\;\hbox{, in}\,H_A\;;\;
\nabla_\Proj^D{\cal E}_B={\cal E}_B\,\gamma_B^D\;\hbox{, in}\;H_B\;,
\end{equation}
where the total potentials are locally given by:
\begin{equation}
\label{totalpot}
\gamma_A^D=\kappa+\mu_A^D\;;\;
\gamma_B^D=\kappa+\mu_B^D\;.
\end{equation}
Here,
\begin{equation}
\label{globpot}
\kappa=\pi_D(\kappa_A)=\pi_D(\kappa_B)
\end{equation}
is a globally defined one-form on $S^2$, while the monopole potentials
\begin{eqnarray}
\mu_A^D& = &{1/2\over 1+\vert\nu_A\vert^2}
\Bigl(\nu_A\dd\nu_A^c-\nu_A^c\dd\nu_A\Bigr)\;,\nonumber\\
\mu_B^D& = &{1/2\over 1+\vert\nu_B\vert^2}
\Bigl(\nu_B\dd\nu_B^c-\nu_B^c\dd\nu_B\Bigr)\;,
\label{localpot}
\end{eqnarray}
are local one-forms, related by a gauge transformation in the overlap
region:
\begin{equation}
\label{ijktransform}
\mu_A^D=\mu_B^D+\Bigl(g^B_A\Bigr)^{-1}\dd g^B_A\;.
\end{equation}
The curvature two-form is also globally defined :
\begin{equation}
\label{field}
\rho=\dd\kappa+\rho_m\;,
\end{equation}
with the monopole field $ \rho_m=\dd\mu_A^D=\dd\mu_B^D$ given by : 
\begin{equation}
\label{monopoolveld}
\rho_m={1\over\Bigl(1+\vert\nu_A\vert^2\Bigr)^2}
\dd\nu_A\wedge\dd\nu_A^c=
{1\over\Bigl(1+\vert\nu_B\vert^2\Bigr)^2}
\dd\nu_B\wedge\dd\nu_B^c\;.
\end{equation}
The integral of the curvature yields the Chern character of the 
projective module:
\begin{equation}
\label{chern}
{\bf ch}\Bigl(\Mod_\Proj\Bigr) ={1\over 2\j\pi}\,\int_{S^2}\rho={1\over 2\j\pi}\,\int_{S^2}\rho_m\;.
\end{equation}
For example, when the homotopy class of $\pi_2(S^2)$ is $[\vec n]=\pm \n$ 
and, provided we choose the representatives in (\ref{positivehomotopy}) 
and (\ref{negativehomotopy}), then 
\begin{equation}
{\bf ch}\Bigl(\Mod_\Proj\Bigr)=\pm
{1\over 2\j\pi}\,\int_{S^2}\;{\n^2\,\vert\zeta_A\vert^{2\n-2}\over
\Bigl(1+\vert\zeta_A\vert^{2\n}\bigr)^2}\;
\dd\zeta_A\wedge\dd\zeta_A^c=\mp\n\;.
\end{equation}
The monopole potential naturally depends on the choice of the representative 
in the relevant homotopy class. With our choice  in (\ref{positivehomotopy}),
they are given in spherical coordinates in $H_A\cap H_B$ by :
\begin {equation}
\label{explicietemonopool}
\mu_A^D=
\mp\j\;\n {({\bf cotg}\theta/2)^{2\n}\over 1+ ({\bf cotg}\theta/2)^{2\n}}
\dd\varphi\;;\;
\mu_B^D=
\pm\j\;\n {({\bf tg}\theta/2)^{2\n}\over 1+ ({\bf tg}\theta/2)^{2\n}}
\dd\varphi\;.
\end{equation}
Both are related by the gauge transformation 
\begin{equation}
\label{explicieteijktransform}
\mu_A^D-\mu_B^D=\mp\j\;\n\,\dd\varphi\;.
\end{equation}
It should be noted that the usual potentials, which are given by
\begin {equation}
\label{gewoonlijkemonopool}
\mu_A^D=
\mp\j\n {({\bf cotg}\theta/2)^{2}\over 1+ ({\bf cotg}\theta/2)^{2}}
\dd\varphi\;;\;
\mu_B^D=
\pm\j\n {({\bf tg}\theta/2)^{2}\over 1+ ({\bf tg}\theta/2)^{2}}
\dd\varphi\;,
\end{equation}
differ from ours by a global differential form.\\
The action for the Yang-Mills potential is, again, obtained from the trace theorem :
\begin{equation}
\label{yangmillsactie}
{\bf S}_{\bf YM}\Bigl[\nabla_P^D\Bigr] = {1\over 2\pi}\,
\int_{S^2}\rho^c\wedge\ast\rho
\end{equation}
The elements $\vert {\bf F}\rangle$ of $\Hi_F$ are given, locally, by 
$\vert{\bf F}\rangle_{\vert A}=E_A\otimes_a\Psi_A$ and\\ 
$\vert{\bf F}\rangle_{\vert B}=E_B\otimes_a\Psi_B$, with 
$E_A$ and $E_B$ (or $\Psi_A$ and $\Psi_B$) related by the passive gauge transformation 
(\ref{gaugetransform4}).
Since ${\bf h}_\Proj(E_A,E_A)=1$, the scalar product of 
$\vert{\bf F}\rangle_{\vert A}$ with 
$\vert{\bf G}\rangle_{\vert A}=E_A\otimes_a\Phi_A$ is :
\begin{equation}
\label{diracskalaarprod}
\langle\!\langle\;\vert {\bf G}\rangle\parallel \vert {\bf F}\rangle\;\rangle\!\rangle=
\langle\Phi_A,\Psi_A\rangle=\int_{S^2}g^{-1}(\Phi_A^c,\Psi_A)\omega\;.
\end{equation}
Here $g^{-1}(\Phi_A^c,\Psi_A)\omega$ may be replaced by 
$\Phi_A^c\wedge(\star\Psi_A)= (\star^\prime\Phi_A^c)\wedge\Psi_A$, 
since forms of degree less than the highest, vanish under the integration symbol. \\
The covariant Dirac operator acts on $\vert{\bf F}\rangle$  as 
$$\DiK(\nabla_\Proj^D)\vert{\bf F}\rangle=
E_A\otimes_a \DiK_A(\gamma_A^D)\Psi_A\;,$$
with
\begin{equation}
\label{diracwerker}
\DiK_A(\gamma_A^D)\Psi_A=\j(\dd-\delta)\Psi_A+\j\gamma_A^D\vee\Psi_A\;.
\end{equation}
Using the selfadjointness of the Dirac operator, the action becomes :
\begin{equation}
\label{diracactie}
{\bf S}_{\bf F}={1\over 2}\int_{S^2}
\Bigl(g^{-1}(\Psi_A^c,\DiK_A\Psi_A)+
g^{-1}((\DiK_A\Psi_A)^c,\Psi_A)\Bigr)\omega\;.
\end{equation}
The Euler-Lagrange equations for the connection are obtained from the total action 
(\ref{actie}) by the variation $\v(\kappa)$ of $\kappa$ in the conection 
$\gamma_A^D=\kappa+\mu_A^D$, so that
$\v(\rho)=\dd\v(\kappa)$ and 
$\v(\DiK_A\Psi_A)=\j\v(\kappa)\vee\Psi_A$.\\
Using the proper formulae of the appendix, it is easy to see that :
\begin{eqnarray}
\v\Bigl({\bf S}_{\bf YM}\Bigr)&=&
{1\over 2\pi}\int_{S^2}
\Bigl(\v(\kappa)^c\wedge\star\delta\rho +
\star^\prime\delta\rho^c\wedge\v(\kappa)\Bigr)\nonumber\\
&=&
{1\over 2\pi}\int_{S^2}
\Bigl(g^{-1}(\v(\kappa)^c,\delta\rho) +
g^{-1}(\delta\rho^c,\v(\kappa))\Bigr)\omega
\label{yangmillsvariatie}
\end{eqnarray}
Under this variation of $\kappa$, the matter action varies as 
$$\v({\bf S}_{\bf F})=
{1\over 2}\int_{S^2}
\Bigl(g^{-1}(\Psi_A^c,\j\v(\kappa)\vee\Psi_A)
+
g^{-1}(-\j\v(\kappa)^c\vee\Psi_A^c,\Psi_A\bigr)\omega\;.$$
Introducing the current of eq.(\ref{stroom}):
\begin{equation}
{\it j}=
g_{\mu\nu}\,g^{-1}(\psi_A^c,\ut^\nu\vee \Psi_A)\;\ut^\mu=
{\it j}^c \;,
\end{equation}
the variation  of the matter action becomes :
\begin{equation}
\label{diracvariatie}
\v({\bf S}_{\bf F})=
{\j\over 2}\int_{S^2}
\Bigl(g^{-1}({\it j}^c,\v(\kappa))-g^{-1}(\v(\kappa)^c,{\it j})\Bigr)\omega\;.
\end{equation}
Combining equations (\ref{yangmillsvariatie}) and (\ref{diracvariatie}) yield
the Euler-Lagrange equations:
\begin{eqnarray}
{\lambda\over 2\pi} \;\delta\rho-{\j\over 2}\,{\it j}&=&0\nonumber\\
{\lambda\over 2\pi} \;\delta\rho^c+{\j\over 2}\,{\it j}^c&=&0\;.
\label{maxwellvergelijking}
\end{eqnarray}
The matter equation results from the variation $\v(\Psi_A)$ in (\ref{diracactie}), 
which yields the covariant Dirac equation of Benn and Tucker \cite{B-T}:
\begin{equation}
\label{diracvergelijking}
\j(\dd-\delta)\Psi_A+\j(\kappa+\mu_A^D)\vee\Psi_A=0\;.
\end{equation}
The system of coupled equations (\ref{maxwellvergelijking}) and 
(\ref{diracvergelijking}) is consistent provided $\delta{\it j}=0$ is 
satisfied and this results from (\ref{stroombehoud}) which holds also for the covariant 
Dirac equation.\\
In the absence of matter, the absolute minimum of ${\bf S}_{\bf YM}$ is 
reached when $\rho=0$, i.e.
\begin{equation}
\label{nulveldvergelijking}
\dd\kappa+\rho_m=0\;.
\end{equation}
Using the Hodge decomposition and ${\bf H}_{deRham}^1(S^2)= 0$, we may write
\begin{equation}
\label{hodgeverdeling}
\kappa=\dd \chi_0+\delta\phi_2\,,
\end{equation}
where $\chi_0$ is a 0-form and $\phi_2$ is a two-form. 
Substitution in (\ref{nulveldvergelijking}) yields
$\dd\delta \phi_2+\rho_m=0$ .
If we put $f=\star\phi_2$, this becomes  
$\dd\star\ua(\dd f) +\rho_m=0 $ or
\begin{equation}
\label{jaye}
\Delta f +\star^{-1}\rho_m=0\;,
\end{equation}
where $\Delta$ is the Laplacian on the sphere.
This is essentially a result of Jayewardena \cite{Jaye} for the 
(classical) Schwinger model on $S^2$.
%
%
\section{Conclusions}
\setcounter{equation}{0}
In this article, we made explicit the procedure of Connes 
and Lott \cite{Con-L} for the algebra of complex valued 
functions on the spere $S^2$, describing the classical 
Schwinger model on the sphere.

Here, the non trivial topological features of the theory show 
up in the projective modules over this algebra with their connections.
For each projective module (i.e. each "sector" of the theory), 
the Lagrangian appears as that of a constrained theory in the sense that the 
monopole field is fixed and  the Euler-Lagrange equations look accordingly.
The Connes-Lott program, restricted here to a commutative algebra,
provides a systematic and consistent way of dealing with topologically 
nontrivial aspects of gauge theories.

A similar treatment of the, still commutative, algebra 
$\diff\otimes ({\bf C}\oplus{\bf C})$ will lead to a generalised
Schwinger model on the sphere and should include
"Higgs" type of phenomena.
Finally, the genuine non-commutative algebra 
of quaternionic-valued functions on $S^4$ should describes 
instantons in the
Connes-Lott framework. 
\section*{Ackowledgements}
We thank Dr.Jos\'e M. Gracia-Bond\'{\i}a for discussions during his visit at UFRJ, Rio de Janeiro,
made posible through the financial support of CLAF(Centro Latino Americano de F\'{\i}sica). 
%
\section*{Appendix}
\renewcommand{\theequation}{A.\arabic{equation}}
\setcounter{equation}{0}
Let $M$ be a Riemannian manifold of dimension $N$ with cotangent bundle  $\tau^*(M) : T^*(M)\rightarrow M$ 
and let $\Lambda^\bullet\bigl(\tau^*(M)\bigr)$ be the exterior product bundle. Its space of sections consists of
the differential  forms $\F^\bullet(M)=\sum_{k=0}^N\F^{(k)}(M)$ which, with the exterior product $\wedge$ and the exterior differential $\dd$, becomes a graded differential algebra
\footnote{We follow the conventions of Kobayashi-Nomizu \cite{K-N} for the Cartan exterior calculus.}.

\nin
The main automorphism $\ua$ of the graded algebra $\bigl\{\F^\bullet(M),\wedge\bigr\}$ is defined by :
\begin{eqnarray}
\ua(\psi\wedge\phi)&=&\ua(\psi)\wedge\ua(\phi)\;,\nonumber\\
\ua(f)&=&f\;,\;f\in\F^{(0)}(M)\;,\;\mbox{and}\nonumber\\ 
\ua(\uxi)&=&-\uxi\;,\;\uxi\in\F^{(1)}(M)\;.\label{mainauto}
\end{eqnarray}
and the main antiautomorphism $\ub$ by :
\begin{eqnarray}
\ub(\psi\wedge\phi)&=&\ub(\phi)\wedge\ub(\psi)\;,\nonumber\\
\ub(f)&=&f\;,\;f\in\F^{(0)}(M)\;,\;\hbox{and}\nonumber\\ 
\ub(\uxi)&=&\uxi\;,\;\uxi\in\F^{(1)}(M)\;.\label{mainantiauto}
\end{eqnarray}
Besides the usual exterior differential $\dd$ and interior product
$\pint(X)$ which are antiderivations on 
$\F^\bullet(M)$ acting from the left, i.e.
\begin{equation}
\dd\bigl(\psi\wedge\phi\bigr)=
(\dd\psi)\wedge\phi+\ua(\psi)\wedge(\dd\phi)\;\mbox{and}
\end{equation}
\begin{equation}
\pint(X)\bigl(\psi\wedge\phi\bigr)=
(\pint(X)\psi)\wedge\phi+\ua(\psi)\wedge(\pint(X)\phi)\;,
\end{equation}
it appears useful to define also a "right" exterior differential and 
interior product by :
\begin{equation}
\dd^D\bigl(\psi\wedge\phi\bigr)=
\psi\wedge(\dd^D\phi)+(\dd^D\psi)\wedge\ua(\phi)\;\mbox{and}
\end{equation}
\begin{equation}
\pint^D(X)\bigl(\psi\wedge\phi\bigr)=
\psi\wedge(\pint^D(X)\phi)+(\pint^D(X)\psi)\wedge\ua(\phi)\;.
\end{equation}
They are related to the usual $\dd$ and $\pint(X)$ by
\footnote{We will also write $\pint(X)\psi=X\rfloor\;\psi$ 
and $\pint^D(X)\psi=\psi\;\rfloor X$}:
\begin{equation}
\dd^D=\dd\circ\ua=-\ua\circ\dd\;,\;\mbox{and}\;\pint^D(X)=
 -\pint(X)\circ\ua=\ua\circ\pint(X)\;.
\end{equation}
The sections of the tangent bundle $\tau(M) : T(M)\rightarrow M$ 
are the vector fields $\X(M)$ on the manifold.
A vector field $X\in\X(M)$ acts on the differential graded algebra 
$\bigl\{\F^\bullet(M),\wedge,\dd\bigr\}$
through the interior product $\pint(X)$ and the Lie derivative 
\begin{equation}
\L(X)=\dd\circ\pint(X)+\pint(X)\circ\dd=
\dd^D\circ\pint^D(X)+\pint^D(X)\circ\dd^D\;.
\end{equation}
Let $\Bigl\{\{\vec e_\mu\},\{\ut^\mu\}\Bigr\}$ be a pair of dual local bases of
the tangent, respectively cotangent, bundle with structure functions given by :
\begin{equation}
[\vec e_\mu,\vec e_\nu]=\vec e_\kappa\;^\kappa C_{\mu\nu}
\quad\mbox{or}\quad
\dd\ut^\kappa=-{1\over 2}\;^\kappa C_{\mu\nu}\,\ut^\mu\wedge\ut^\nu\;.
\end{equation}
The metric on the manifold defines a scalar product on $\F^\bullet(M)$ 
with values in $C^\infty(M)$:
\[
g^{-1}:\F^\bullet(M)\times \F^\bullet(M)\rightarrow C^\infty(M):
(\psi,\phi)\rightarrow g^{-1}\bigl(\psi,\phi\bigr)\;,
\]
which is $C^\infty(M)$-bilinear, symmetric and such that forms of 
different order are orthogonal.
Let 
$\psi=\psi_{\alpha_1\cdots\alpha_k}\,\ut^{\alpha_1}
\wedge \cdots\wedge\ut^{\alpha_k}\;,\;
\phi=\phi_{\beta_1\cdots\beta_k}\,\ut^{\beta_1}
\wedge\cdots\wedge\ut^{\beta_k}$, then
\begin{equation}
\label{scalarproduct}
g^{-1}\bigl(\psi,\phi\bigr)=
k!\;\psi_{\alpha_1\cdots\alpha_k}\;\phi_{\beta_1\cdots\beta_k}\;
g^{\alpha_1\beta_1}\cdots g^{\alpha_k\beta_k}
\end{equation}
This product has the following properties
\begin{equation}
\label{prodprop}
g^{-1}\bigl(\uxi\wedge\chi,\phi\bigr)=
g^{-1}\bigl(\chi,\tilde{\uxi}\rfloor\;\phi\bigr)\quad,\quad
g^{-1}\bigl(\chi\wedge\uxi,\phi\bigr)=
g^{-1}\bigl(\chi,\phi\;\rfloor\tilde{\uxi}\bigr)\;,
\end{equation}
where $\tilde{\uxi}$ is the vector field defined by 
\begin{equation}
\label{indlift}
\tilde{\uxi}\rfloor\;\ueta=\ueta\;\rfloor\tilde{\uxi}=
g^{-1}\bigl(\ueta,\uxi\bigr)\;.
\end{equation}
The Hodge duals $\star$ and $\star\p$ are the $C^\infty(M)$-linear mappings :
\begin{eqnarray*}
\star_{\,k}& :\F^{(k)}(M)\rightarrow\F^{(N-k)}(M)&:
\phi\rightarrow\star_{\,k}\phi\;,\\
\star\p_{\,k}& :\F^{(k)}(M)\rightarrow\F^{(N-k)}(M)&:
\psi\rightarrow\star\p_{\,k}\psi
\end{eqnarray*}
given by :
\begin{equation}
\label{Hodge}
\psi\wedge(\star_{\,k}\phi)=(\star\p_{\,k}\psi)\wedge\phi=g^{-1}\bigl(\psi,\phi\bigr)\,\omega\;.
\end{equation}
They are related by 
\begin{equation}
\star_{\,k}=(-1)^{k(N-k)}\star\p_{\,k}
\end{equation}
and on inhomogeneous forms the action of $\star$ and $\star\p$
is straightforward.

\nin
The above properties (\ref{prodprop}) of $g^{-1}$ imply that
\begin{equation}
\label{Hodgeprop1}
\star\bigl(\phi\wedge\uxi\bigr)=\tilde\uxi\rfloor\,(\star\phi)\quad,\quad
\star\p\bigl(\ueta\wedge\psi\bigr)=(\star\p\psi)\,\rfloor\tilde\ueta\;.
\end{equation}
Since $\;\star 1=\star\p 1=\omega\;$, the repeated application 
of (\ref{Hodgeprop1}) leads to :
\begin{eqnarray}
\label{Hodgeform}
\star\bigl(\uxi^1\wedge\cdots\uxi^k\bigr)&=&
\tilde\uxi^k\rfloor\,\cdots\tilde\uxi^1\rfloor\,\omega\;,\nonumber\\
\star\p\bigl(\uxi^k\wedge\cdots\uxi^1\bigr)&=&
\omega\,\rfloor\tilde\uxi^1\cdots\,\rfloor\tilde\uxi^k\;.
\end{eqnarray}
The inverse of the Hodge star-operators are given by :
\begin{equation}
\star\p_{\,N-k}\circ\,\star_{\,k}=
\star_{\,N-k}\circ\star\p_{\,k}=\hbox{Sign}(det\,g)\,{\bf Id}_{\,k}\;.
\end{equation}
Other useful properties are :
\begin{equation}\star\circ\,\ua\,=\,(-1)^N\,\ua\circ\star
\;,\;
\star\p\circ\,\ua\,=\,(-1)^N\,\ua\circ\star\p\;,
\end{equation}
\begin{equation}
g^{-1}\bigl(\star\,\psi,\phi\bigr)=g^{-1}\bigl(\psi,\star\p\phi\bigr)\;,
\end{equation}
\begin{equation}
\uxi\wedge\bigl(\star\phi\bigr)=\star\bigl(\phi\,\rfloor\tilde\uxi\bigr)\;,\;
\bigl(\star\p\psi\bigr)\wedge\ueta=
\star\p\bigl(\tilde\ueta\rfloor\,\psi\bigr)\;.
\end{equation}
The complexification of the exterior bundle yields complex-valued 
differential forms ${{\cal F}^\bullet(M)}^{\bf C}$ with a naturally 
defined operation of complex conjugation $\psi\rightarrow\psi^c$ and 
an involution $\psi\rightarrow\psi^{\dagger}$ related by 
$\psi^c=\ub\bigl(\psi^\dagger\bigr)$.\\
On ${{\cal F}^\bullet(M)}^{\bf C}$  a $\diff^{\bf C}$-sesquilinear form 
is given by :
\begin{equation}
\label{sesquilin}
h^{-1}\bigl(\psi,\phi\bigr)=g^{-1}\bigl(\psi^c,\phi\bigr)\,.
\end{equation}
When $M$ is compact\footnote{If $M$ were not compact, the scalar
product can be defined for differential forms of compact support
or for square integrable forms with the Riemannian invariant measure.}, it 
defines a Hermitian scalar product in ${{\cal F}^\bullet(M)}^{\bf C}$ :
\begin{equation}
\label{hermitprod}
\langle\psi\;\vert\;\phi\rangle=
\int_{M}\,h^{-1}\bigl(\psi,\phi\bigr)\,\omega=
\int_{M}\,\psi^c\wedge(\star\phi)=
\int_{M}\,(\star\p\psi^c)\wedge\phi\;.
\end{equation}
The completion of ${{\cal F}^\bullet(M)}^{\bf C}$ with 
this product (\ref{hermitprod}) yields 
a Hilbert space $\Hi(M)$.
The adjoint operators of $\dd$ and $\dd^D$ with respect to this scalar
product are the codifferentials $\delta$ and $\delta^D$ obtained as follows.
\begin{eqnarray*}
\langle\dd\psi\vert\phi\rangle&=&
\int_{M}\,(\dd\psi^c)\wedge(\star\phi)\\
&=&
\int_{M}\,\dd\bigl(\psi^c\wedge(\star\phi)\bigr)\,-\,
\int_{M}\,\ua(\psi^c)\wedge\dd(\star\phi)\\
&=&
\int_{M}\,\psi^c\wedge\bigl(\star(\star^{-1}(\dd(\star(\ua(\phi)))))\bigr)\;\\
&=&
\langle\psi\vert\delta\phi\rangle\;,
\end{eqnarray*}
so that
\begin{equation}
\label{codif1}
\delta=\star^{-1}\circ\dd\circ\star\circ\ua\;=\;
(-1)^{N+1}\,{\star\p}^{-1}\circ\dd\circ\star\p\circ\ua\;.
\end{equation}
In the same way
\[
\langle\psi\vert\dd^D\phi\rangle=
\int_{M}\,(\star\p\psi^c)\wedge(\dd^D\phi)=\cdots=
\langle\delta^D\psi\vert\phi\rangle\;,
\]
and
\begin{equation}
\label{codif2}
\delta^D={\star\p}^{-1}\circ\dd^D\circ\star\p\circ\ua\;=\;
(-1)^{N+1}\,\star^{-1}\circ\dd^D\circ\star\circ\ua\;.
\end{equation}
Both codifferentials are related by :
\begin{equation}
\delta^D=\ua\circ\delta=-\delta\circ\ua\;.
\end{equation}
Just as the exterior algebra $\Lambda^\bullet(\tau^*)$ is obtained as 
the quotient of the tensor algebra $\bigotimes^\bullet(\tau^*)$ by the 
two-sided ideal ${\cal I}_{ext}$ generated by elements of the form
$\{\uxi\otimes\ueta\,+\,\ueta\otimes\uxi\}$ : 
\[
\{\Lambda^\bullet(\tau^*)=
\bigotimes^\bullet(\tau^*)/{\cal I}_{ext},\wedge\}\;,
\]
the Clifford algebra\footnote{A detailed account on Clifford algebra and the usual 
Dirac operator with special emphasis to applications in noncommutative geometry, 
can be found in the lecture notes 
"{\bf CLIFFORD GEOMETRY: A Seminar}" 
by J.C.V\'{a}rilly and J.M.Gracia-Bond\'{\i}a of the University of Costa Rica.} 
is obtained as the quotient by the ideal 
${\cal I}_{Cliff}$ generated by elements of the form 
$\{\uxi\otimes\ueta\,+\,\ueta\otimes\uxi\,-\,2g^{-1}(\uxi,\ueta)\,{bf 1}\}$,
\[
\{C\ell^{(\pm)}(\tau^*)=
\bigotimes^\bullet(\tau^*)/{\cal I}_{Cliff},\vee\}\;.
\]
Since ${\cal I}_{ext}$ is generated by homogeneous elements, the {\bf Z}-grading of 
$\bigotimes^\bullet(\tau^*)$ persists in $\Lambda^\bullet(\tau^*)$, while 
${\cal I}_{Cliff}$ being generated by inhomogeneous but even elements of $\bigotimes^\bullet(\tau^*)$, 
only a ${\bf Z}_2$ grading survives in $C\ell^{(\pm)}(\tau^*)$.
\newpage
\nin
As vector spaces both $\Lambda^\bullet(\tau^*)$ and $C\ell^{(\pm)}(\tau^*)$ are
isomorphic so that they can be considered as a single vector space with two 
different products $\wedge$ and $\vee$, yielding the so-called 
Atiyah-K\H{a}hler algebra. This algebraic construction can be done in the 
cotangent space of each point of the manifold $M$, yielding an 
Atiyah-K\H{a}hler algebra-bundle with its space of sections 
$\;\{\F^\bullet(M),\wedge,\vee\}$.
The relation between the two products was given by 
K\H{a}hler(see\cite{Graf}) and, in our notation, reads   
\begin{equation}
\label{AtKae}
\psi\vee\phi=
\sum_{k=1}^N\;{1\over k!} g^{\alpha_1\beta_1}\ldots
g^{\alpha_k\beta_k}
\bigl(\bigl(\ldots\bigl(\psi\,\rfloor\vec{e_{\alpha_1}}\bigr)\ldots
\bigr)\rfloor\vec{e_{\alpha_k}}\bigr)\;\wedge
\bigl(\vec{e_{\beta_k}}\rfloor\bigl(\cdots\bigl(
\vec{e_{\beta_1}}\rfloor\phi\bigr)\cdots\bigr)\bigr)\;.
\end{equation}
In particular 
\begin{equation}
\label{Clifprod1}
\uxi\vee\phi=\uxi\wedge\phi+\tilde\uxi\rfloor\,\phi\;,
\end{equation}
and
\begin{equation}
\label{Clifprod2}
\psi\vee\ueta=\psi\wedge\ueta+\psi\,\rfloor\tilde\ueta\;.
\end{equation}
Further useful formulae are :
\[
g^{-1}\bigl(\xi\vee\psi,\phi\bigl)=g^{-1}\bigl(\psi,\xi\vee\phi\bigl)\;,\;
g^{-1}\bigl(\psi,\phi\vee\ueta\bigl)=g^{-1}\bigl(\psi\vee\ueta,\phi\bigl)\;.
\]
Also
\[
\star\bigl(\psi\vee\ueta\bigr)=\ueta\vee\bigl(\star\psi\bigr)\;,\;
\star\p\bigl(\uxi\vee\phi\bigr)=\bigl(\star\p\phi\bigr)\vee\uxi\;,
\]
which imply :
\begin{eqnarray*}
\star\bigl(\ueta^1\vee\ldots\vee\ueta^k\bigr)&=&
\bigl(\ueta^k\vee\ldots\vee\ueta^1\bigr)\vee\omega\;,\\
\star\p\bigl(\uxi^1\vee\ldots\vee\uxi^k\bigr)&=&
\omega\vee\bigl(\uxi^k\vee\ldots\vee\uxi^1\bigr)\;,
\end{eqnarray*}
\nin
or, more generally :
\begin{equation}
\star\phi=\ub\bigl(\phi\bigr)\vee\omega\;,\;
\star\p\psi=\omega\vee\ub\bigl(\psi\bigr)\;.
\end{equation}
The exterior differentials and codifferentials can be written in terms
of the Levi-Civita covariant derivative as follows :
\begin{eqnarray}
\dd\psi=\ut^\mu\wedge\bigl(\nabla_{\vec e_\mu}\psi\bigr)
&\;\mbox{and}\;&
\delta\psi=-\tilde\ut^\mu\rfloor\,\bigl(\nabla_{\vec e_\mu}\psi\bigr)\;;
\label{difcodif}\\
\dd^D\psi=\bigl(\nabla_{\vec e_\mu}\psi\bigr)\wedge\ut^\mu\;,
&\;\mbox{and}\;&
\delta^D\psi=-\bigl(\nabla_{\vec e_\mu}\psi\bigr)\,\rfloor\tilde\ut^\mu\;.
\label{difcodifD}
\end{eqnarray}
The Hermitian K\H{a}hler-Dirac operators are defined by :
\begin{eqnarray}
\DiK\psi&=\j\bigl(\dd-\delta\bigr)\psi&
=\j\ut^\mu\vee\bigl(\nabla_{\vec e_\mu}\psi\bigr)\;,\label{DiKae1}\\
\DiK^D\psi&=\j\bigl(\dd^D-\delta^D\bigr)\psi&
=\j\bigl(\nabla_{\vec e_\mu}\psi\bigr)\vee\ut^\mu\;.\label{DiKae2}
\end{eqnarray}
A local current can be constructed as follows :
\begin{eqnarray*}
g^{-1}\bigl({1\over\j}\DiK\psi^c,\phi\bigr)
&=&g^{-1}\bigl(\ut^\mu\vee(\nabla_{\vec e_\mu}\psi^c),\phi\bigr)
=g^{-1}\bigl(\nabla_{\vec e_\mu}\psi^c,\ut^\mu\vee\phi\bigr) 
\\
&=&\vec e_\mu\Bigl(g^{-1}\bigl(\psi^c,\ut^\mu\vee\phi\bigr)\Bigr)
-
g^{-1}\bigl(\psi^c,\nabla_{\vec e_\mu}(\ut^\mu\vee\phi)\bigr)
\\
&=&
\vec e_\mu\Bigl(g^{-1}\bigl(\psi^c,\ut^\mu\vee\phi\bigr)\Bigr)
-
g^{-1}\bigl(\psi^c,(\nabla_{\vec e_\mu}\ut^\mu)\vee\phi)\bigr)\\
& & \;-\,
g^{-1}\bigl(\psi^c,\ut^\mu\vee(\nabla_{\vec e_\mu}\phi)\bigr)\;. 
\end{eqnarray*}
So that the current 
\begin{equation}
\label{current1}
\J^\mu=g^{-1}\bigl(\psi^c,\ut^\mu\vee\phi\bigr)
\end{equation}
obeys 
\begin{equation}
\vec e_\mu(\J^\mu)+\Gamma^\mu_{\mu\,\nu}\J^\nu=
g^{-1}\Bigl(({1\over\j}\DiK\psi)^c,\phi\Bigr)+
g^{-1}\Bigl(\psi^c,{1\over\j}\DiK\phi\Bigr)
\nonumber
\end{equation}
and is covariantly conserved if $\psi$ and $\phi$ obey the Dirac-K\H{a}hler
equation\\ $\DiK\psi=0=\DiK\phi$.\\
The current density 
\begin{equation}
\label{current2}
\JD^\mu=g^{-1}\bigl(\psi^c,\ut^\mu\vee\phi\bigr)\,\sqrt{det\,g}
\end{equation}
is then divergence free :
\begin{equation}
\label{conserv}
{\bf div}\JD=\vec e_\mu(\JD^\mu)+C^\mu_{\mu\,\nu}\JD^\nu=0\;.
\end{equation}
The current one-form
\begin{equation}
\label{stroom}
\underline{ j}= g_{\mu\nu}\;g^{-1}(\psi^c,\ut^\nu\vee\psi)\ut^\mu\;,
\end{equation}
obeys 
\begin{equation}
\label{stroombehoud}
\delta \underline{j}=0
\end{equation}
and is dual to the (N-1)-form of Benn and Tucker \cite{B-T} 
which is closed when the Dirac-K\H{a}hler equation is satisfied.

\end{document}